\documentclass[12pt]{article}

\usepackage{jheppub, bm, wrapfig,float,array}
\usepackage[utf8]{inputenc}
\numberwithin{equation}{section}
\renewcommand\afterTocRuleSpace{\bigskip}

\usepackage{amsfonts}
\usepackage{amsmath}
\usepackage{color}
\usepackage{graphicx}

\usepackage{tikz}
\usetikzlibrary{arrows}
\usetikzlibrary{shapes.geometric,calc,arrows, positioning,shapes.misc,decorations.markings}
\tikzset{
  big arrow/.style={
    decoration={markings,mark=at position 1 with {\arrow[scale=2,#1]{>}}},
    postaction={decorate},
    shorten >=0.4pt},
  big arrow/.default=black}

\newcommand{\bea}{\begin{eqnarray}}
\newcommand{\eea}{\end{eqnarray}}
\newcommand{\be}{\begin{equation}}
\newcommand{\ee}{\end{equation}}
\newcommand{\bit}{\begin{itemize}}
\newcommand{\eit}{\end{itemize}}
\newcommand{\ben}{\begin{enumerate}}
\newcommand{\een}{\end{enumerate}}

\newcommand{\half}{\frac{1}{2}}

\newcommand{\cN}{\mathcal{N}}

\newcommand{\fg}{\mathfrak{g}}
\newcommand{\fh}{\mathfrak{h}}
\newcommand{\su}{\mathfrak{su}}
\renewcommand{\sp}{\mathfrak{sp}}
\newcommand{\so}{\mathfrak{so}}

\title{Revisiting the classifications of $6d$ SCFTs and LSTs}

\author{Lakshya Bhardwaj\footnote{lbhardwaj at fas.harvard.edu}}

\affiliation{Department of Physics, Harvard University, Cambridge, MA 02138, USA}

\abstract{Gauge-theoretic anomaly cancellation predicts the existence of many $6d$ SCFTs and little string theories (LSTs) that have not been given a string theory construction so far. In this paper, we provide an explicit construction of all such ``missing'' $6d$ SCFTs and LSTs by using the frozen phase of F-theory. We conjecture that the full set of $6d$ SCFTs and LSTs is obtained by combining the set of theories constructed in this paper with the set of theories that have been constructed in earlier literature using the unfrozen phase of F-theory. Along the way, we demonstrate that there exist SCFTs that do not descend from LSTs via an RG flow.}

\begin{document}
\nocite{*}
\maketitle

\section{Introduction and Conclusions} \label{intro}
$6d$ SCFTs and little string theories (LSTs) have been at the focal point of many recent developments in quantum field theory and string theory \cite{Gaiotto:2014lca,DelZotto:2014hpa,Heckman:2014qba,Sakai:2014hsa,Ohmori:2014pca,Ohmori:2014kda,Intriligator:2014eaa,Haghighat:2014vxa,DelZotto:2014fia,Karndumri:2014sha,Apruzzi:2015zna,DelZotto:2015isa,Karndumri:2015rsa,Gaiotto:2015usa,Ohmori:2015pua,Hohenegger:2015cba,Kim:2015jba,Gadde:2015tra,DelZotto:2015rca,Heckman:2015ola,Hayashi:2015fsa,Yonekura:2015ksa,Cordova:2015fha,Passias:2015gya,Heckman:2015axa,Ohmori:2015pia,Zafrir:2015rga,Ohmori:2015tka,Hayashi:2015zka,Kim:2015fxa,Bertolini:2015bwa,Anderson:2015cqy,Heckman:2016ssk,Apruzzi:2016rny,Font:2016odl,Morrison:2016nrt,Morrison:2016djb,Johnson:2016qar,Yun:2016yzw,Haghighat:2016jjf,Kim:2016foj,Shimizu:2016lbw,DelZotto:2016pvm,Heckman:2016xdl,Apruzzi:2016nfr,Razamat:2016dpl,Lawrie:2016axq,Bobev:2016phc,Mekareeya:2016yal,Gu:2017ccq,Kim:2017xan,Yankielowicz:2017xkf,Hayashi:2017jze,DelZotto:2017pti,Bah:2017wxp,Chacaltana:2017boe,Couzens:2017way,Haghighat:2017vch,Chang:2017xmr,Jefferson:2017ahm,Choi:2017vtd,Bastian:2017jje,Mekareeya:2017jgc,Mekareeya:2017sqh,Dibitetto:2017klx,Apruzzi:2017iqe,Heckman:2017uxe,Font:2017cya,Kim:2017toz,Hassler:2017arf,Merkx:2017jey,Bastian:2017ary,Bourton:2017pee,Apruzzi:2017nck,DelZotto:2017mee,Mayrhofer:2017nwp,Nazzal:2018brc,Kim:2018gjo,Jefferson:2018irk,Anderson:2018heq,Bah:2018gwc,Apruzzi:2018oge,Lee:2018ihr,Dierigl:2018nlv,DelZotto:2018tcj,Haghighat:2018dwe,Cvetic:2018xaq,Heckman:2018pqx,Bastian:2018fba,Gu:2018gmy,Tian:2018icz,Frey:2018vpw,Haghighat:2018gqf,Cordova:2018eba,Bhardwaj:2018vuu,Gukov:2018iiq,Apruzzi:2018nre,Ohmori:2018ona,Kim:2018bpg,Hanany:2018vph,Kim:2018lfo,Razamat:2018gbu,Zafrir:2018hkr,Duan:2018sqe,Buican:2016hpb,Bobev:2015kza,Hohenegger:2015btj,Hohenegger:2016eqy,Hohenegger:2016yuv,Haouzi:2016yyg,Bastian:2017ing,Haouzi:2017vec,Kachru:2018van,Bastian:2018jlf,Filippas:2019puw,Naseer:2018cpj,Zhu:2017ysu,Bhardwaj:2018yhy,Merkx:2019bmm,Nunez:2018ags}. Many of these developments were inspired by the classifications of these theories carried out in \cite{Heckman:2015bfa,Heckman:2013pva,Bhardwaj:2015xxa,Bhardwaj:2015oru}. These classifications have taken two different starting points. On one hand are the classifications of \cite{Heckman:2015bfa,Heckman:2013pva,Bhardwaj:2015oru} which study all the $6d$ SCFTs and LSTs which can be constructed by compactifying F-theory on an elliptically fibered Calabi-Yau threefold. These classifications are incomplete, because as pointed out in \cite{Bhardwaj:2018jgp}, the F-theory compactifications considered by \cite{Heckman:2015bfa,Heckman:2013pva,Bhardwaj:2015oru} do not include frozen singularities. On the other hand is the classification of \cite{Bhardwaj:2015xxa} which studies all the consistent\footnote{The consistency conditions are based on a  version of Green-Schwarz mechanism of anomaly cancellation in the six-dimensional context, which was first discussed in \cite{Sagnotti:1992qw}.} $6d$ supersymmetric gauge theories that can arise as low energy theories on the tensor branch of a $6d$ SCFT or LST, and conjectures that the corresponding $6d$ SCFTs and LSTs exist. Such a classification is incomplete because there exist $6d$ SCFTs and LSTs that are not described purely by a $6d$ supersymmetric gauge theory on their tensor branch.

To compare the two classifications, one can compare the set of theories obtained in \cite{Bhardwaj:2015xxa} to the subset of those theories in \cite{Heckman:2015bfa,Heckman:2013pva,Bhardwaj:2015oru} that are described purely by a gauge theory on their tensor branch. One finds that some of the theories obtained in \cite{Bhardwaj:2015xxa} are missing from \cite{Heckman:2015bfa,Heckman:2013pva,Bhardwaj:2015oru}. We can divide such theories into two types:
\ben
\item First of all, there are theories which are known to have a field-theoretic inconsistency even though they solve the consistency conditions imposed in \cite{Bhardwaj:2015xxa}. See \cite{Ohmori:2015pia} for an example. 
\item Second, there are theories that involve sub-quivers that cannot be constructed in F-theory without frozen singularities, but admit a construction once we allow frozen singularities in F-theory. See \cite{Bhardwaj:2018jgp} for a construction of some of these sub-quivers. It is these theories that will be the main topic of discussion in this paper. It is interesting to note that some, but not all, of these theories are known to admit a brane construction in massive type IIA string theory\footnote{See \cite{Brunner:1997gk} for initial work on Hanany-Witten-like brane constructions of six-dimensional theories.} for around 20 years now \cite{Brunner:1997gf,Hanany:1997gh,Hanany:1999sj}.
\een

This paper is organized as follows. In Section \ref{MT}, we list down all of the possible missing theories that involve sub-quivers that cannot be constructed in F-theory without frozen singularities\footnote{We emphasize that our list also includes those theories that contain non-gauge-theoretic factors like E-string and $\cN=(2,0)$ theory. This is unlike \cite{Bhardwaj:2015xxa} where the discussion was entirely restricted to gauge theories.}. We continue in Section \ref{r} with a brief discussion about the reasons for the omission of such theories from the unfrozen phase of F-theory. Then, in Section \ref{If}, we introduce new constructions of various sub-quivers that we need to construct the theories listed in Section \ref{MT}. Finally, in Sections \ref{CS} and \ref{CL}, we go on to explicitly show how each theory listed in Section \ref{MT} can be constructed by compactifying F-theory on an elliptically fibered Calabi-Yau threefold involving frozen singularities.

We conjecture that the full list of $6d$ SCFTs and LSTs is obtained by combining the classification of this paper with the earlier classification of \cite{Heckman:2015bfa,Bhardwaj:2015oru}. Our conjecture stems from the fact that this combined classification exhausts all the possible tensor branches that can be obtained by putting together gauge theories with known non-gauge theories like the E-string theory and $\mathsf{A}_1$ $(2,0)$ theory. We caution that there is a small set of theories whose F-theory construction was proposed in \cite{Heckman:2015bfa,Bhardwaj:2015oru} but
a closer look in \cite{Merkx:2017jey} (see also \cite{Morrison:2016djb,Bertolini:2015bwa}) revealed an inconsistency in the proposed constructions of those theories. It would be worthwhile to investigate whether such theories can be given a consistent construction in the frozen phase of F-theory. We leave this as an interesting problem for future work.

As a by-product of our work, we demonstrate the existence of SCFTs that do not descend from LSTs via an RG flow. See (\ref{S5}), (\ref{S9}) and (\ref{S10}) for examples of such theories and (\ref{CS5}), (\ref{CS9}), (\ref{CS10}) for their F-theory constructions. Such SCFTs were earlier expected to be inconsistent in \cite{Bhardwaj:2015oru} because as shown there almost all SCFTs do admit a LST completion. As shown in this paper, this expectation is not correct.

\section{Missing theories}\label{MT}
We start in Section \ref{SQ} by listing down all the sub-quivers appearing in \cite{Bhardwaj:2015xxa} but not admitting a construction in the unfrozen phase of F-theory. We then list down all the possible LSTs and SCFTs containing these sub-quivers\footnote{We slightly enlarge the extent of the classification of \cite{Bhardwaj:2015xxa} by allowing some non-gauge-theoretic factors to appear in the low energy theory on the tensor branch in the form of formal gauge algebras $\sp(0)$ and $\su(1)$.} in Sections \ref{ML} and \ref{MS} respectively. In compiling our list, we discard those theories which involve certain sub-quivers known to have a field theoretic inconsistency \cite{Ohmori:2015pia}.

\subsection{Missing sub-quivers}\label{SQ}
\bit
\item \be  \label{MI1}
\begin{tikzpicture}
\node (v1) at (-0.45,0.9) {$\su(n)$};
\begin{scope}[shift={(-1.5,0)}]
\node (v0) at (-0.45,0.9) {$\mathsf{S}^2$};
\end{scope}
\draw  (v0) edge (v1);
\end{tikzpicture}
\ee
which denotes a hyper in two-index symmetric representation $\mathsf{S}^2$ of $\su(n)$.
\item \be
\begin{tikzpicture}
\node (v1) at (-0.45,0.9) {$\su(n)$};
\begin{scope}[shift={(2,0)}]
\node (v2) at (-0.45,0.9) {$\so(m)$};
\end{scope}
\draw  (v1) edge (v2);
\end{tikzpicture}
\ee
where the edge denotes a hyper in bifundamental of $\su\oplus\so$.
\item \be
\begin{tikzpicture}
\node (v1) at (-0.45,0.9) {$\su(4)$};
\begin{scope}[shift={(2,0)}]
\node (v2) at (-0.45,0.9) {$\so(7)$};
\end{scope}
\draw  (v1) edge (v2);
\node at (0.9,1) {\tiny{$\mathsf{S}$}};
\end{tikzpicture}
\ee
where the edge decorated by $\mathsf{S}$ on one side denotes a hyper in $\text{fundamental}\otimes\text{spinor}$ of $\su\oplus\so$.
\item \be
\begin{tikzpicture}
\node (v1) at (-0.45,0.9) {$\su(4)$};
\begin{scope}[shift={(1.8,0)}]
\node (v2) at (-0.45,0.9) {$\fg_2$};
\end{scope}
\draw  (v1) edge (v2);
\end{tikzpicture}
\ee
where the edge denotes a hyper in $\text{fundamental}\otimes\mathbf{7}$ of $\su\oplus\fg_2$.
\item \be
\begin{tikzpicture}
\node (v1) at (-0.5,0.45) {$\so(n_1)$};
\begin{scope}[shift={(1.8,0)}]
\node (v2) at (-0.5,0.45) {$\sp(n)$};
\end{scope}
\begin{scope}[shift={(3.6,0)}]
\node (v3) at (-0.5,0.45) {$\so(n_3)$};
\end{scope}
\begin{scope}[shift={(1.8,1.2)}]
\node (v4) at (-0.5,0.45) {$\so(n_2)$};
\end{scope}
\draw  (v1) -- (v2);
\draw  (v2) -- (v3);
\draw  (v2) edge (v4);
\end{tikzpicture}
\ee
where the edge between $\sp(n)$ and $\so(n_i)$ denotes a half-hyper in bifundamental of $\sp(n)\oplus\so(n_i)$.
\item \be
\begin{tikzpicture}
\node (v1) at (-0.5,0.45) {$\so(n_1)$};
\begin{scope}[shift={(1.8,0)}]
\node (v2) at (-0.5,0.45) {$\sp(n)$};
\end{scope}
\begin{scope}[shift={(3.6,0)}]
\node (v3) at (-0.5,0.45) {$\su(n_3)$};
\end{scope}
\begin{scope}[shift={(1.8,1.2)}]
\node (v4) at (-0.5,0.45) {$\so(n_2)$};
\end{scope}
\draw  (v1) -- (v2);
\draw  (v2) -- (v3);
\draw  (v2) edge (v4);
\end{tikzpicture}
\ee
where the edge between $\sp(n)$ and $\su(n_3)$ denotes a hyper in bifundamental of $\sp(n)\oplus\su(n_3)$.
\item \be
\begin{tikzpicture}
\node (v1) at (-0.5,0.45) {$\so(n_1)$};
\begin{scope}[shift={(1.8,0)}]
\node (v2) at (-0.5,0.45) {$\sp(4)$};
\end{scope}
\begin{scope}[shift={(3.6,0)}]
\node (v3) at (-0.5,0.45) {$\so(7)$};
\end{scope}
\begin{scope}[shift={(1.8,1.2)}]
\node (v4) at (-0.5,0.45) {$\so(n_2)$};
\end{scope}
\draw  (v1) -- (v2);
\draw  (v2) -- (v3);
\draw  (v2) edge (v4);
\node at (2.45,0.55) {\tiny{$\mathsf{S}$}};
\end{tikzpicture}
\ee
where the edge decorated by $\mathsf{S}$ on one side denotes a half-hyper in $\text{fundamental}\otimes\text{spinor}$ of $\sp\oplus\so$.
\item \be
\begin{tikzpicture}
\node (v1) at (-0.5,0.45) {$\so(n_1)$};
\begin{scope}[shift={(1.8,0)}]
\node (v2) at (-0.5,0.45) {$\sp(4)$};
\end{scope}
\begin{scope}[shift={(3.6,0)}]
\node (v3) at (-0.5,0.45) {$\fg_2$};
\end{scope}
\begin{scope}[shift={(1.8,1.2)}]
\node (v4) at (-0.5,0.45) {$\so(n_2)$};
\end{scope}
\draw  (v1) -- (v2);
\draw  (v2) -- (v3);
\draw  (v2) edge (v4);
\end{tikzpicture}
\ee
where the edge between $\sp$ and $\fg_2$ denotes a half-hyper in $\text{fundamental}\otimes\mathbf{7}$ of $\sp\oplus\fg_2$.
\item \be
\begin{tikzpicture}
\node (v1) at (-0.5,0.45) {$\so(7)$};
\begin{scope}[shift={(1.8,0)}]
\node (v2) at (-0.5,0.45) {$\sp(2)$};
\end{scope}
\begin{scope}[shift={(3.6,0)}]
\node (v3) at (-0.5,0.45) {$\so(7)$};
\end{scope}
\begin{scope}[shift={(1.8,1.2)}]
\node (v4) at (-0.5,0.45) {$\so(7)$};
\end{scope}
\draw  (v1) -- (v2);
\draw  (v2) -- (v3);
\draw  (v2) edge (v4);
\node at (2.45,0.55) {\tiny{$\mathsf{S}$}};
\node at (0.2,0.55) {\tiny{$\mathsf{S}$}};
\node at (1.175,1.175) {\tiny{$\mathsf{S}$}};
\end{tikzpicture}
\ee
\eit

\subsection{Missing LSTs} \label{ML}
Let us first list down all the possible LSTs carrying the sub-quivers listed in Section \ref{SQ}:
\bit
\item \be \label{L1}
\begin{tikzpicture}
\node (v1) at (-0.45,0.9) {$\su(n_0)$};
\begin{scope}[shift={(-1.5,0)}]
\node (v0) at (-0.45,0.9) {$\mathsf{S}^2$};
\end{scope}
\begin{scope}[shift={(1.8,0)}]
\node (v2) at (-0.45,0.9) {$\su(n_1)$};
\end{scope}
\begin{scope}[shift={(3.1,0)}]
\node (v3) at (-0.45,0.9) {$\cdots$};
\end{scope}
\begin{scope}[shift={(4.4,0)}]
\node (v4) at (-0.45,0.9) {$\su(n_k)$};
\end{scope}
\begin{scope}[shift={(6.2,0)}]
\node (v5) at (-0.45,0.9) {$\sp(m)$};
\end{scope}
\draw  (v0) edge (v1);
\draw  (v1) edge (v2);
\draw  (v2) edge (v3);
\draw  (v3) edge (v4);
\draw  (v4) edge (v5);
\end{tikzpicture}
\ee
where all the edges except the leftmost one denote a hyper in bifundamental. Here $n_i=2m+8+8(k-i)$ with $m\ge0$ and $k\ge0$. The case $m=0$ corresponds to an E-string theory at the rightmost end of the quiver.

Its construction is given in (\ref{CL1}).
\item \be \label{L2}
\begin{tikzpicture}
\node (v1) at (-0.45,0.9) {$\su(n_0)$};
\begin{scope}[shift={(-1.5,0)}]
\node (v0) at (-0.45,0.9) {$\mathsf{S}^2$};
\end{scope}
\begin{scope}[shift={(1.8,0)}]
\node (v2) at (-0.45,0.9) {$\su(n_1)$};
\end{scope}
\begin{scope}[shift={(3.1,0)}]
\node (v3) at (-0.45,0.9) {$\cdots$};
\end{scope}
\begin{scope}[shift={(4.4,0)}]
\node (v4) at (-0.45,0.9) {$\su(n_j)$};
\end{scope}
\begin{scope}[shift={(5.7,0)}]
\node (v5) at (-0.45,0.9) {$\cdots$};
\end{scope}
\begin{scope}[shift={(7,0)}]
\node (v6) at (-0.45,0.9) {$\su(n_k)$};
\end{scope}
\begin{scope}[shift={(8.8,0)}]
\node (v7) at (-0.45,0.9) {$\sp(0)$};
\end{scope}
\begin{scope}[shift={(4.4,1)}]
\node (w) at (-0.45,0.9) {$\mathsf{F}$};
\end{scope}
\draw  (v0) edge (v1);
\draw  (v1) edge (v2);
\draw  (v2) edge (v3);
\draw  (v3) edge (v4);
\draw  (v4) edge (v5);
\draw  (v5) edge (v6);
\draw  (v6) edge (v7);
\draw  (w) edge (v4);
\end{tikzpicture}
\ee
where the edge between $\su(n_j)$ and $\mathsf{F}$ denotes a hyper in the fundamental representation $\mathsf{F}$ of $\su(n_j)$. Here $n_i=9+9(k-i)$ for $j\le i\le k$ and $n_i=9+9(k-j)+8(j-i)$ for $0\le i\le j$ with $0\le j\le k$ and $k\ge0$. $\sp(0)$ is a shorthand for E-string which allows a neighboring $\su(n\le9)$. Since these theories involve an E-string, they don't appear in \cite{Bhardwaj:2015xxa} but can be obtained by a mild extension of the rules considered there.

Its construction is given in (\ref{CL1}).
\item \be \label{L3}
\begin{tikzpicture}
\node (v1) at (-0.45,0.9) {$\su(n_0)$};
\begin{scope}[shift={(-1.5,0)}]
\node (v0) at (-0.45,0.9) {$\mathsf{S}^2$};
\end{scope}
\begin{scope}[shift={(1.8,0)}]
\node (v2) at (-0.45,0.9) {$\su(n_1)$};
\end{scope}
\begin{scope}[shift={(3.1,0)}]
\node (v3) at (-0.45,0.9) {$\cdots$};
\end{scope}
\begin{scope}[shift={(4.4,0)}]
\node (v4) at (-0.45,0.9) {$\su(n_k)$};
\end{scope}
\begin{scope}[shift={(6.2,0)}]
\node (v5) at (-0.45,0.9) {$\su(m)$};
\end{scope}
\begin{scope}[shift={(7.7,0)}]
\node (v6) at (-0.45,0.9) {$\mathsf{\Lambda}^2$};
\end{scope}
\draw  (v0) edge (v1);
\draw  (v1) edge (v2);
\draw  (v2) edge (v3);
\draw  (v3) edge (v4);
\draw  (v4) edge (v5);
\draw  (v5) edge (v6);
\end{tikzpicture}
\ee
where the rightmost edge denotes a hyper in two-index antisymmetric representation $\mathsf{\Lambda}^2$ of $\su(m)$. Here $n_i=m+8+8(k-i)$ with $m\ge2$ and $k\ge0$.

Its construction is given in (\ref{CL2}).
\item \be \label{L4}
\begin{tikzpicture}
\node (v1) at (-0.45,0.9) {$\su(n_0)$};
\begin{scope}[shift={(-1.5,0)}]
\node (v0) at (-0.45,0.9) {$\mathsf{S}^2$};
\end{scope}
\begin{scope}[shift={(1.8,0)}]
\node (v2) at (-0.45,0.9) {$\su(n_1)$};
\end{scope}
\begin{scope}[shift={(3.1,0)}]
\node (v3) at (-0.45,0.9) {$\cdots$};
\end{scope}
\begin{scope}[shift={(4.4,0)}]
\node (v4) at (-0.45,0.9) {$\su(n_j)$};
\end{scope}
\begin{scope}[shift={(5.7,0)}]
\node (v5) at (-0.45,0.9) {$\cdots$};
\end{scope}
\begin{scope}[shift={(7,0)}]
\node (v6) at (-0.45,0.9) {$\su(n_k)$};
\end{scope}
\begin{scope}[shift={(8.6,0)}]
\node (v7) at (-0.45,0.9) {$\half\mathsf{\Lambda}^3$};
\end{scope}
\begin{scope}[shift={(4.4,1)}]
\node (w) at (-0.45,0.9) {$\mathsf{F}$};
\end{scope}
\draw  (v0) edge (v1);
\draw  (v1) edge (v2);
\draw  (v2) edge (v3);
\draw  (v3) edge (v4);
\draw  (v4) edge (v5);
\draw  (v5) edge (v6);
\draw  (v6) edge (v7);
\draw  (w) edge (v4);
\end{tikzpicture}
\ee
where the rightmost edge denotes a half-hyper in three-index antisymmetric representation $\mathsf{\Lambda}^3$ of $\su(n_k)$. Here $n_i=6+9(k-i)$ for $j\le i\le k$ and $n_i=6+9(k-j)+8(j-i)$ for $0\le i\le j$ with $0\le j\le k$ and $k\ge1$.

Its construction is given in (\ref{CL3}).
\item \be \label{L5}
\begin{tikzpicture}
\node (v1) at (-0.45,0.9) {$\su(n_1)$};
\begin{scope}[shift={(-1.8,0)}]
\node (v0) at (-0.45,0.9) {$\so(n_0)$};
\end{scope}
\begin{scope}[shift={(1.8,0)}]
\node (v2) at (-0.45,0.9) {$\su(n_2)$};
\end{scope}
\begin{scope}[shift={(3.1,0)}]
\node (v3) at (-0.45,0.9) {$\cdots$};
\end{scope}
\begin{scope}[shift={(4.4,0)}]
\node (v4) at (-0.45,0.9) {$\su(n_k)$};
\end{scope}
\begin{scope}[shift={(6.2,0)}]
\node (v5) at (-0.45,0.9) {$\sp(m)$};
\end{scope}
\draw  (v0) edge (v1);
\draw  (v1) edge (v2);
\draw  (v2) edge (v3);
\draw  (v3) edge (v4);
\draw  (v4) edge (v5);
\end{tikzpicture}
\ee
Here $n_i=2m+8+8(k-i)$ with $m\ge0$ and $k\ge1$.

Its construction is given in (\ref{CL4}).
\item \be \label{L6}
\begin{tikzpicture}
\node (v1) at (-0.45,0.9) {$\su(n_1)$};
\begin{scope}[shift={(-1.8,0)}]
\node (v0) at (-0.45,0.9) {$\so(n_0)$};
\end{scope}
\begin{scope}[shift={(1.8,0)}]
\node (v2) at (-0.45,0.9) {$\su(n_2)$};
\end{scope}
\begin{scope}[shift={(3.1,0)}]
\node (v3) at (-0.45,0.9) {$\cdots$};
\end{scope}
\begin{scope}[shift={(4.4,0)}]
\node (v4) at (-0.45,0.9) {$\su(n_j)$};
\end{scope}
\begin{scope}[shift={(5.7,0)}]
\node (v5) at (-0.45,0.9) {$\cdots$};
\end{scope}
\begin{scope}[shift={(7,0)}]
\node (v6) at (-0.45,0.9) {$\su(n_k)$};
\end{scope}
\begin{scope}[shift={(8.8,0)}]
\node (v7) at (-0.45,0.9) {$\sp(0)$};
\end{scope}
\begin{scope}[shift={(4.4,1)}]
\node (w) at (-0.45,0.9) {$\mathsf{F}$};
\end{scope}
\draw  (v0) edge (v1);
\draw  (v1) edge (v2);
\draw  (v2) edge (v3);
\draw  (v3) edge (v4);
\draw  (v4) edge (v5);
\draw  (v5) edge (v6);
\draw  (v6) edge (v7);
\draw  (w) edge (v4);
\end{tikzpicture}
\ee
Here $n_i=9+9(k-i)$ for $j\le i\le k$ and $n_i=9+9(k-j)+8(j-i)$ for $0\le i\le j$ with $1\le j\le k$ and $k\ge1$.

Its construction is given in (\ref{CL4}).

For $j=0$, we obtain
\be \label{L7}
\begin{tikzpicture}
\node (v1) at (-0.45,0.9) {$\su(n_1)$};
\begin{scope}[shift={(-1.8,0)}]
\node (v0) at (-0.45,0.9) {$\so(n_0)$};
\end{scope}
\begin{scope}[shift={(1.8,0)}]
\node (v2) at (-0.45,0.9) {$\su(n_2)$};
\end{scope}
\begin{scope}[shift={(3.1,0)}]
\node (v3) at (-0.45,0.9) {$\cdots$};
\end{scope}
\begin{scope}[shift={(4.4,0)}]
\node (v4) at (-0.45,0.9) {$\su(n_k)$};
\end{scope}
\begin{scope}[shift={(6.2,0)}]
\node (v5) at (-0.45,0.9) {$\sp(0)$};
\end{scope}
\begin{scope}[shift={(-1.8,1)}]
\node (w) at (-0.45,0.9) {$\mathsf{F}$};
\end{scope}
\draw  (v0) edge (v1);
\draw  (v1) edge (v2);
\draw  (v2) edge (v3);
\draw  (v3) edge (v4);
\draw  (v4) edge (v5);
\draw  (w) edge (v0);
\end{tikzpicture}
\ee
where the edge between $\so(n_0)$ and $\mathsf{F}$ denotes a hyper in the fundamental representation $\mathsf{F}$ of $\so(n_0)$. Here $n_i=9+9(k-i)$ with $k\ge1$.

Its construction is given in (\ref{CL4}).
\item \be \label{L8}
\begin{tikzpicture}
\node (v1) at (-0.45,0.9) {$\su(n_1)$};
\begin{scope}[shift={(-1.8,0)}]
\node (v0) at (-0.45,0.9) {$\so(n_0)$};
\end{scope}
\begin{scope}[shift={(1.8,0)}]
\node (v2) at (-0.45,0.9) {$\su(n_2)$};
\end{scope}
\begin{scope}[shift={(3.1,0)}]
\node (v3) at (-0.45,0.9) {$\cdots$};
\end{scope}
\begin{scope}[shift={(4.4,0)}]
\node (v4) at (-0.45,0.9) {$\su(n_k)$};
\end{scope}
\begin{scope}[shift={(6.2,0)}]
\node (v5) at (-0.45,0.9) {$\su(m)$};
\end{scope}
\begin{scope}[shift={(7.7,0)}]
\node (v6) at (-0.45,0.9) {$\mathsf{\Lambda}^2$};
\end{scope}
\draw  (v0) edge (v1);
\draw  (v1) edge (v2);
\draw  (v2) edge (v3);
\draw  (v3) edge (v4);
\draw  (v4) edge (v5);
\draw  (v5) edge (v6);
\end{tikzpicture}
\ee
Here $n_i=m+8+8(k-i)$ with $m\ge2$ and $k\ge1$.

Its construction is given in (\ref{CL5}).
\item \be \label{L9}
\begin{tikzpicture}
\node (v1) at (-0.45,0.9) {$\su(n_1)$};
\begin{scope}[shift={(-1.8,0)}]
\node (v0) at (-0.45,0.9) {$\so(n_0)$};
\end{scope}
\begin{scope}[shift={(1.8,0)}]
\node (v2) at (-0.45,0.9) {$\su(n_2)$};
\end{scope}
\begin{scope}[shift={(3.1,0)}]
\node (v3) at (-0.45,0.9) {$\cdots$};
\end{scope}
\begin{scope}[shift={(4.4,0)}]
\node (v4) at (-0.45,0.9) {$\su(n_j)$};
\end{scope}
\begin{scope}[shift={(5.7,0)}]
\node (v5) at (-0.45,0.9) {$\cdots$};
\end{scope}
\begin{scope}[shift={(7,0)}]
\node (v6) at (-0.45,0.9) {$\su(n_k)$};
\end{scope}
\begin{scope}[shift={(8.6,0)}]
\node (v7) at (-0.45,0.9) {$\half\mathsf{\Lambda}^3$};
\end{scope}
\begin{scope}[shift={(4.4,1)}]
\node (w) at (-0.45,0.9) {$\mathsf{F}$};
\end{scope}
\draw  (v0) edge (v1);
\draw  (v1) edge (v2);
\draw  (v2) edge (v3);
\draw  (v3) edge (v4);
\draw  (v4) edge (v5);
\draw  (v5) edge (v6);
\draw  (v6) edge (v7);
\draw  (w) edge (v4);
\end{tikzpicture}
\ee
Here $n_i=6+9(k-i)$ for $j\le i\le k$ and $n_i=6+9(k-j)+8(j-i)$ for $1\le i\le j$ with $1\le j\le k$ and $k\ge2$.

Its construction is given in (\ref{CL6}).

For $j=0$, we obtain
\be \label{L10}
\begin{tikzpicture}
\node (v1) at (-0.45,0.9) {$\su(n_1)$};
\begin{scope}[shift={(-1.8,0)}]
\node (v0) at (-0.45,0.9) {$\so(n_0)$};
\end{scope}
\begin{scope}[shift={(1.8,0)}]
\node (v2) at (-0.45,0.9) {$\su(n_2)$};
\end{scope}
\begin{scope}[shift={(3.1,0)}]
\node (v3) at (-0.45,0.9) {$\cdots$};
\end{scope}
\begin{scope}[shift={(4.4,0)}]
\node (v4) at (-0.45,0.9) {$\su(n_k)$};
\end{scope}
\begin{scope}[shift={(6,0)}]
\node (v5) at (-0.45,0.9) {$\half\mathsf{\Lambda}^3$};
\end{scope}
\begin{scope}[shift={(-1.8,1)}]
\node (w) at (-0.45,0.9) {$\mathsf{F}$};
\end{scope}
\draw  (v0) edge (v1);
\draw  (v1) edge (v2);
\draw  (v2) edge (v3);
\draw  (v3) edge (v4);
\draw  (v4) edge (v5);
\draw  (w) edge (v0);
\end{tikzpicture}
\ee
Here $n_i=6+9(k-i)$ with $k\ge2$.

Its construction is given in (\ref{CL6}).
\item \be \label{L11}
\begin{tikzpicture}
\node (v1) at (-0.45,0.9) {$\sp(n_2)$};
\begin{scope}[shift={(-3.6,0)}]
\node (v-1) at (-0.45,0.9) {$\su(n_0)$};
\end{scope}
\begin{scope}[shift={(-1.8,0)}]
\node (v0) at (-0.45,0.9) {$\so(n_1)$};
\end{scope}
\begin{scope}[shift={(1.8,0)}]
\node (v2) at (-0.45,0.9) {$\so(n_3)$};
\end{scope}
\begin{scope}[shift={(3.1,0)}]
\node (v3) at (-0.45,0.9) {$\cdots$};
\end{scope}
\begin{scope}[shift={(4.5,0)}]
\node (v4) at (-0.45,0.9) {$\sp(n_{2k})$};
\end{scope}
\begin{scope}[shift={(6.3,0)}]
\node (v5) at (-0.45,0.9) {$\su(m)$};
\end{scope}
\draw  (v0) edge (v1);
\draw  (v1) edge (v2);
\draw  (v2) edge (v3);
\draw  (v3) edge (v4);
\draw  (v4) edge (v5);
\draw  (v-1) edge (v0);
\end{tikzpicture}
\ee
where the dots denote an alternating $\sp-\so$ chain. We remind the reader that edges between $\so$ and $\sp$ correspond to a half-hyper rather than a full hyper in bifundamental. Here $n_{2i+1}=2n_{2i}=2m+16(k-i)$ with $m\ge2$ and $k\ge1$.

Its construction is given in (\ref{CL7}).
\item \be \label{L12}
\begin{tikzpicture}
\node (v1) at (-0.45,0.9) {$\so(n_1)$};
\begin{scope}[shift={(-1.8,0)}]
\node (v0) at (-0.45,0.9) {$\su(n_0)$};
\end{scope}
\begin{scope}[shift={(1.8,0)}]
\node (v2) at (-0.45,0.9) {$\sp(n_2)$};
\end{scope}
\begin{scope}[shift={(3.1,0)}]
\node (v3) at (-0.45,0.9) {$\cdots$};
\end{scope}
\begin{scope}[shift={(4.5,0)}]
\node (v4) at (-0.45,0.9) {$\sp(n_{2j})$};
\end{scope}
\begin{scope}[shift={(5.9,0)}]
\node (v5) at (-0.45,0.9) {$\cdots$};
\end{scope}
\begin{scope}[shift={(7.3,0)}]
\node (v6) at (-0.45,0.9) {$\sp(n_{2k})$};
\end{scope}
\begin{scope}[shift={(9.4,0)}]
\node (v7) at (-0.45,0.9) {$\su(1)$};
\end{scope}
\begin{scope}[shift={(4.5,1.1)}]
\node (w) at (-0.45,0.9) {$\half\mathsf{F}$};
\end{scope}
\draw  (v0) edge (v1);
\draw  (v1) edge (v2);
\draw  (v2) edge (v3);
\draw  (v3) edge (v4);
\draw  (v4) edge (v5);
\draw  (v5) edge (v6);
\draw  (v6) edge (v7);
\draw  (w) edge (v4);
\node at (7.8,1.1) {\tiny{$\half\mathsf{F}$}};
\end{tikzpicture}
\ee
where the dots denote alternating $\sp-\so$ chains and the edge between $\sp(n_{2j})$ and $\half\mathsf{F}$ denotes a half-hyper in fundamental representation $\mathsf{F}$ of $\sp(n_{2j})$. $\su(1)$ at the rightmost node indicates an unpaired tensor corresponding to $\mathsf{A}_1$ $\cN=(2,0)$ theory. The decoration by $\half\mathsf{F}$ on top of rightmost edge indicates that a half-hyper in fundamental of $\sp(n_{2k})=\sp(1)$ has to be trapped there for the edge between $\sp(n_{2k})=\sp(1)$ and $\su(1)$ to be consistent\footnote{The existence of this trapped $\half\mathsf{F}$ can be understood if one views the $\mathsf{A}_1$ $\cN=(2,0)$ theory in the $\cN=(1,0)$ language. The $\cN=(2,0)$ R-symmetry is $\so(5)$ whose $\so(4)$ subalgebra  decomposes into $\su(2)$ $\cN=(1,0)$ R-symmetry plus an $\su(2)=\sp(1)$ flavor symmetry. The $\cN=(2,0)$ tensor multiplet decomposes into a $\cN=(1,0)$ tensor multiplet plus a $\cN=(1,0)$ hypermultiplet such that the hypermultiplet transforms as $\half\mathsf{F}$ under the flavor $\sp(1)$. This flavor $\sp(1)$ is gauged in (\ref{L12}) by the gauge algebra $\sp(n_{2k})=\sp(1)$.}. This half-hyper is unlike the half-hyper attached to $\sp(n_{2j})$ because the latter can move around as we change $j$ but the former must remain attached to $\sp(n_{2k})=\sp(1)$. Here $n_{2i+1}+1=2n_{2i}=2+18(k-i)$ for $j\le i\le k$ and $n_{2i+1}=2n_{2i}=2+18(k-j)+16(j-i)$ for $0\le i\le j-1$ with $1\le j\le k$ and $k\ge1$.

Its construction is given in (\ref{CL8}) and (\ref{CL8'}).

For $j=0$, we obtain
\be \label{L12'}

\ee
where the dots denote an alternating $\sp-\so$ chain. Here $n_{2i+1}+1=2n_{2i}=2+18(k-i)$ with $k\ge1$.

Its construction is given in (\ref{CL8}).
\item \be \label{L13}
%
\ee
where the dots denote an alternating $\sp-\so$ chain. Here $n_{2i+1}=2n_{2i}=16+16(k-i)$ with $k\ge1$. The sub-quiver 
\be
%
\ee
formed by the two rightmost nodes denotes a rank two E-string theory.

Its construction is given in (\ref{CL9}).

For $k=0$, we obtain
\be \label{L14}
%
\ee

Its construction is given in (\ref{CL10}).
\item \be \label{L15}
%
\ee
where the dots denote an alternating $\sp-\so$ chain and the rightmost edge denotes a hyper in spinor representation $\mathsf{S}$ of $\so(n_{2k+1})$. Here $n_{2i+1}=2n_{2i}=12+16(k-i)$ with $k\ge1$.

Its construction is given in (\ref{CL11}).

For $k=0$, we obtain
\be \label{L16}
%
\ee

Its construction is given in (\ref{CL12}).
\item \be \label{L17}
%
\ee
where the dots denote alternating $\sp-\so$ chains and the rightmost edge denotes a half-hyper in spinor representation $\mathsf{S}$ of $\so(n_{2k+1})$. Here $n_{2i+1}+1=2n_{2i}=14+18(k-i)$ for $j\le i\le k$ and $n_{2i+1}=2n_{2i}=14+18(k-j)+16(j-i)$ for $0\le i\le j-1$ with $1\le j\le k$ and $k\ge1$.

Its construction is given in (\ref{CL13}).

For $j=0$, we obtain
\be \label{L17'}
%
\ee
where the dots denote an alternating $\sp-\so$ chain. Here $n_{2i+1}+1=2n_{2i}=14+18(k-i)$ with $k\ge1$.

Its construction is given in (\ref{CL13}).

For $k=0$, we obtain
\be \label{L18}
%
\ee

Its construction is given in (\ref{CL14}).
\item \be \label{L19}
%
\ee
where the dots denote an alternating $\sp-\so$ chain. Here $n_{2i+1}=2n_{2i}=8+16(k-i)$ with $k\ge1$.

Its construction is given in (\ref{CL15}).

For $k=0$, we obtain
\be \label{L20}
%
\ee
where the dots denote alternating $\sp-\so$ chains. Here $n_{2i+1}+1=2n_{2i}=8+18(k-i)$ for $j\le i\le k$ and $n_{2i+1}=2n_{2i}=8+18(k-j)+16(j-i)$ for $0\le i\le j-1$ with $1\le j\le k$ and $k\ge1$.

Its construction is given in (\ref{CL17}).

For $j=0$, we obtain
\be \label{L22}
%
\ee
where the dots denote an alternating $\sp-\so$ chain. Here $n_{2i+1}+1=2n_{2i}=8+18(k-i)$ with $k\ge1$.

Its construction is given in (\ref{CL17}).

For $k=0$, we obtain
\be \label{L23}
%
\ee
where the dots denote an alternating $\sp-\so$ chain. Here $n_{2i+1}=2n_{2i}=4m+16+16(k-i)$ with $m\ge0$ and $k\ge1$. The case $m=0$ gives rise to two E-string factors at the right end of the quiver.

Its construction is given in (\ref{CL20}).

For $k=0$, we obtain
\be \label{L25}
%
\ee
Here $n_{1}=2n_{0}=4m+16$ with $m\ge0$.

Its construction is given in (\ref{CL21}).
\item \be \label{L26}
%
\ee
where the dots denote an alternating $\sp-\so$ chain. Here $n_{2i+1}=2n_{2i}=2m+16(k-i)$ and $n=m+8+8k$ with $m\ge2$ and $k\ge1$.

Its construction is given in (\ref{CL22}).
\item \be \label{L27}
%
\ee
where the dots denote alternating $\sp-\so$ chains. Here $n_{2i+1}+1=2n_{2i}=2+18(k-i)$ for $j\le i\le k$, $n_{2i+1}=2n_{2i}=2+18(k-j)+16(j-i)$ for $0\le i\le j-1$, and $n=9+9k-j$ with $0\le j\le k$ and $k\ge1$.

Its construction is given in (\ref{CL23}) and (\ref{CL23'}).
\item \be \label{L28}
%
\ee
where the dots denote an alternating $\sp-\so$ chain. Here $n_{2i+1}+1=2n_{2i}=2+18(k-i)$ and $n=9+9k$ with $k\ge1$.

Its construction is given in (\ref{CL24}).
\item \be \label{L29}
%
\ee
where the dots denote an alternating $\sp-\so$ chain. Here $n_{2i+1}=2n_{2i}=16(k-i)$ and $n=8+8k$ with $k\ge1$. The two rightmost nodes gives rise to a rank two E-string factor in the low energy theory.

Its construction is given in (\ref{CL25}).
\item \be \label{L30}
%
\ee
where the dots denote an alternating $\sp-\so$ chain. Here $n_{2i+1}=2n_{2i}=12+16(k-i)$ and $n=14+8k$ with $k\ge0$.

Its construction is given in (\ref{CL26}).
\item \be \label{L31}
%
\ee
where the dots denote alternating $\sp-\so$ chains. Here $n_{2i+1}+1=2n_{2i}=14+18(k-i)$ for $j\le i\le k$, $n_{2i+1}=2n_{2i}=14+18(k-j)+16(j-i)$ for $0\le i\le j-1$ and $n=15+9k-j$ with $0\le j\le k$ and $k\ge0$.

Its construction is given in (\ref{CL27}).
\item \be \label{L32}
%
\ee
where the dots denote an alternating $\sp-\so$ chain. Here $n_{2i+1}+1=2n_{2i}=14+18(k-i)$ and $n=15+9k$ with $k\ge0$.

Its construction is given in (\ref{CL28}).
\item \be \label{L33}
%
\ee
where the dots denote an alternating $\sp-\so$ chain. Here $n_{2i+1}=2n_{2i}=8+16(k-i)$ and $n=12+8k$ with $k\ge1$.

Its construction is given in (\ref{CL29}).

For $k=0$, we obtain
\be \label{L34}
%
\ee
where the dots denote alternating $\sp-\so$ chains. Here $n_{2i+1}+1=2n_{2i}=8+18(k-i)$ for $j\le i\le k$, $n_{2i+1}=2n_{2i}=8+18(k-j)+16(j-i)$ for $0\le i\le j-1$ and $n=12+9k-j$ with $0\le j\le k$ and $k\ge1$.

Its construction is given in (\ref{CL31}).

For $k=0$, we obtain
\be \label{L36}
%
\ee
where the dots denote an alternating $\sp-\so$ chain. Here $n_{2i+1}+1=2n_{2i}=8+18(k-i)$ and $n=12+9k$ with $k\ge1$.

Its construction is given in (\ref{CL34}).

For $k=0$, we obtain
\be \label{L38}
%
\ee
where the dots denote an alternating $\sp-\so$ chain. Here $n_{2i+1}=2n_{2i}=4m+16+16(k-i)$ and $n=2m+16+8k$ with $m\ge0$ and $k\ge0$.

Its construction is given in (\ref{CL36}).
\item \be \label{L40}
%
\ee
with $m\ge0$.

Its construction is given in (\ref{CL37}).
\item \be \label{L42}
%
\ee

Its construction is given in (\ref{CL38}).
\eit

\subsection{Missing SCFTs} \label{MS}
Let us now list down all the possible SCFTs carrying the sub-quivers listed in Section \ref{SQ}. Our list below will contain SCFTs that do not have an LST parent. These SCFTs are (\ref{S5}), (\ref{S9}) and (\ref{S10}).
\bit
\item \be \label{S1}
%
\ee
where the edge between $\su(n_i)$ and $m_i\mathsf{F}$ denotes $m_i$ hypers in fundamental of $\su(n_i)$. Here $m_0=n_0-8-n_1$ and $m_i=2n_i-n_{i-1}-n_{i+1}$ for $1\le i\le k$ with $n_{k+1}:=0$ and $k\ge0$.

Its construction is given in (\ref{CS1}).
\item \be \label{S2}
%
\ee
where the edge between $\so(n_0)$ and $m_0\mathsf{F}$ denotes $m_0$ hypers in vector of $\so(n_0)$. Here $m_0=n_0-8-n_1$ and $m_i=2n_i-n_{i-1}-n_{i+1}$ for $1\le i\le k$ with $n_{k+1}:=0$ and $k\ge1$.

Its construction is given in (\ref{CS2}).
\item \be  \label{S3}
%
\ee
where the dots denote an alternating $\sp-\so$ chain and the edge between $\sp(n_{2i})$ and $m_{2i}\mathsf{F}$ denotes $m_{2i}$ hypers in fundamental of $\sp(n_{2i})$. Here $m_0=2n_0-n_1$, $m_{2i-1}=n_{2i-1}-8-n_{2i-2}-n_{2i}$ and $m_{2i}=2n_{2i}+8-\frac{n_{2i-1}}{2}-\frac{n_{2i+1}}{2}$ for $1\le i\le k$ with $n_{2k+1}:=0$ and $k\ge1$. Here $n_{2k}$ can be zero, in which case we obtain an E-string factor at the right end of the quiver.

Its construction is given in (\ref{CS3}).
\item \be  \label{S4}
%
\ee
where the dots denote an alternating $\sp-\so$ chain. Here $m=2n-n_0$, $m_0=n_0-8-n_1-n$, $m_{2i}=n_{2i}-8-n_{2i-1}-n_{2i}$ and $m_{2i+1}=2n_{2i+1}+8-\frac{n_{2i}}{2}-\frac{n_{2i+2}}{2}$ for $1\le i\le k$ with $n_{2k+1}:=0$ and $k\ge0$.

Its construction is given in (\ref{CS4}).
\item \be \label{S5}
%
\ee
Here $m_0=2n_0-n_1$, $m_1=2n_1-n_0-n_2$, $m_2=n_2-8-n_1-n_3$ and $m_3=2n_3+8-\frac{n_2}{2}$.

Its construction is given in (\ref{CS5}). It was suspected in \cite{Bhardwaj:2015oru} that this theory is probably not consistent since there is no LST from which it can be obtained by decoupling a tensor multiplet. Our construction in (\ref{CS5}) demonstrates that this suspicion is not correct, and shows that there exist SCFTs that cannot be obtained via an RG flow starting from a LST.
\item \be \label{S6}
%
\ee
where the dots denote an alternating $\sp-\so$ chain. Here $m_0=n_0-8-n_2$, $m_1=n_1-8-n_2$, $m_2=2n_2+8-\frac{n_0}{2}-\frac{n_1}{2}-\frac{n_3}{2}$, $m_{2i-1}=n_{2i-1}-8-n_{2i}-n_{2i-2}$ and $m_{2i}=2n_{2i}+8-\frac{n_{2i-1}}{2}-\frac{n_{2i+1}}{2}$ for $2\le i\le k$ with $n_{2k+1}:=0$ and $k\ge2$. Here $n_{2k}$ can be zero, in which case we obtain an E-string factor at the right end of the quiver.

Its construction is given in (\ref{CS6}).
\item \be \label{S7}
%
\ee
where the dots denote an alternating $\sp-\so$ chain. Here $m_0=n_0-8-n_2$, $m_1=n_1-8-n_2$, $m_2=2n_2+8-\frac{n_0}{2}-\frac{n_1}{2}-\frac{n_3}{2}$, $m_{2i-1}=n_{2i-1}-8-n_{2i}-n_{2i-2}$, $m_{2i}=2n_{2i}+8-\frac{n_{2i-1}}{2}-\frac{n_{2i+1}}{2}$ and $m_{2k+1}=n_{2k+1}-8-n_{2k}$ for $2\le i\le k$ with $k\ge1$.

Its construction is given in (\ref{CS7}).
\item \be \label{S8}
%
\ee
Here $m_0=2n_0+8-\frac{n_1}{2}$, $m_1=n_1-8-n_0-n_2$, $m_2=2n_2+8-\frac{n_1}{2}-\frac{n_3}{2}-\frac{n_5}{2}$, $m_3=n_3-8-n_4-n_2$, $m_4=2n_4+8-\frac{n_3}{2}$ and $m_5=n_5-8-n_2$.

Its construction is given in (\ref{CS8}).
\item \be \label{S9}
%
\ee
Here $m_0=2n_0+8-\frac{n_1}{2}$, $m_1=n_1-8-n_0-n_2$, $m_2=2n_2+8-\frac{n_1}{2}-\frac{n_3}{2}-\frac{n_6}{2}$, $m_3=n_3-8-n_4-n_2$, $m_4=2n_4+8-\frac{n_3}{2}-\frac{n_5}{2}$, $m_5=n_5-8-n_4$ and $m_6=n_6-8-n_2$.

Its construction is given in (\ref{CS9}). Like (\ref{S5}), this theory is an example of an SCFT that cannot be obtained from an LST via an RG flow.
\item \be \label{S10}
%
\ee
Here $m_0=2n_0+8-\frac{n_1}{2}$, $m_1=n_1-8-n_0-n_2$, $m_2=2n_2+8-\frac{n_1}{2}-\frac{n_3}{2}-\frac{n_7}{2}$, $m_3=n_3-8-n_4-n_2$, $m_4=2n_4+8-\frac{n_3}{2}-\frac{n_5}{2}$, $m_5=n_5-8-n_4-n_6$, $m_6=2n_6+8-\frac{n_5}{2}$ and $m_7=n_7-8-n_2$.

Its construction is given in (\ref{CS10}). Like (\ref{S5}) and (\ref{S9}), this theory is another example of an SCFT that cannot be obtained from an LST via an RG flow.
\eit

\section{$6d$ SCFTs and LSTs from the frozen phase}\label{FTC}
\subsection{Reasons for missing theories}\label{r}
We now recall the reasons due to which the theories listed in Sections \ref{ML} and \ref{MS} do not admit a construction in the unfrozen phase of F-theory. These theories can be divided into three types.

The first type of theories involve an $\su(n)$ gauge algebra with a hyper in $\mathsf{S}^2$ and $n-8$ hypers in $\mathsf{F}$. For such a theory to admit a construction in the unfrozen phase of F-theory, the $\su(n)$ must arise on a curve $C$ in the base $B$ of the F-theory compactification such that:
\ben
\item The arithmetic genus of $C$ must be one.
\item The self-intersection of $C$ in $B$ must be $-1$.
\een
It was shown in Appendix B of \cite{Heckman:2013pva} that the order of vanishing of $(f,g)$ appearing in the Weierstrass model on such a curve $C$ is at least $(4,6)$. Such a large order of vanishing of $(f,g)$ on a curve in $B$ is considered to be unphysical. Hence, no such theory can be constructed in the unfrozen phase of F-theory.

The second type of theories involve an $\su(m\ge4)$ gauge algebra with $2m$ hypers in $\mathsf{F}$ such that a subset of those hypers transform in a representation $\mathsf{R}$ of another gauge algebra which is either $\so(n)$ or $\fg_2$. For such a theory to admit a construction in the unfrozen phase of F-theory, the following conditions must be satisfied:
\ben
\item The $\su(m)$ must arise on a curve $C$ and $\so(n)$ or $\fg_2$ must arise on a curve $D$ such that $C\cdot D\neq0$.
\item The $\so(n)$ or $\fg_2$ algebra must arise from an $\text{I}^*_{p}$ singularity over $D$. 
\item Since $m\ge4$, $\su(m)$ must arise from an $\text{I}_m$ singularity over $C$. 
\item $C$ must have genus zero and self-intersection $-2$.
\een
Now, an $\text{I}_m$ singularity over such a $C$ cannot consistently intersect an $\text{I}^*_p$ singularity. Thus, no such theory can be constructed in the unfrozen phase of F-theory.

The third type of theories involve an $\sp(m\ge2)$ gauge algebra with $2m+8$ hypers in $\mathsf{F}$ such that three subsets of those hypers transform respectively in representation $\mathsf{R}_1$, $\mathsf{R}_2$ and $\mathsf{R}_3$ of other gauge algebras $\fh_1$, $\fh_2$ and $\fh_3$ such that each $\fh_i$ is either an $\so$ algebra or a $\fg_2$ algebra. For such a theory to admit a construction in the unfrozen phase of F-theory, the following conditions must be satisfied:
\ben
\item The $\sp(m)$ must arise on a curve $C$ and $\fh_i$ must arise on a curve $D_i$ such that $C\cdot D_i\neq0$ for each $i$.
\item The $\fh_i$ must arise from an $\text{I}^*_{p_i}$ singularity over $D$.
\item Since $m\ge2$, $\sp(m)$ must arise from a non-split $\text{I}_{2m}$ singularity over $C$.
\item $C$ must have genus zero and self-intersection $-1$.
\een
Now, an $\text{I}_{2m}$ singularity over such a $C$ cannot consistently intersect three singularities $\text{I}^*_{p_i}$. Thus, no such theory can be constructed in the unfrozen phase of F-theory.

\subsection{Ingredients from the frozen phase}\label{If}
\subsubsection{New constructions of old ingredients}
The frozen phase provides us with novel constructions of some gauge-theoretic ingredients that already admit a construction in the unfrozen phase. We will use the following constructions in this paper:
\ben
\item $\sp(m)$ gauge algebra with $(2m+8)\mathsf{F}$ can be constructed in the frozen phase by a curve\footnote{All of the curves considered in this paper have genus zero.} $C$ of self-intersection $-4$ carrying an $\hat{\text{I}}^*_{m+4}$ singularity where, following the notation of \cite{Bhardwaj:2018jgp}, we add a hat on top of an $\text{I}^*_n$ singularity if it carries an algebra of $\sp$ type\footnote{Notice that $n\ge4$ for an $\hat{\text{I}}^*_{n}$ singularity.} rather than $\so$ type. In type IIB language, an $\hat{\text{I}}^*_{m+4}$ singularity corresponds to a stack of $m$ D7 branes on top of an O7$^+$ plane\footnote{In our notation, a superscript $+$ denotes an O7 plane of positive RR charge and a superscript $-$ denotes an O7 plane of negative RR charge.}.

There are a total of $4m+16$ zeroes of the residual discriminant $\tilde\Delta_C$ on $C$. Each zero carries a $\half\mathsf{F}$ of $\sp(m)$ leading to a total of $(2m+8)\mathsf{F}$ of $\sp(m)$. If all the points on $C$ where $\tilde\Delta_C$ vanishes have even multiplicity of zeroes, then the $\hat{\text{I}}^*_{m+4}$ singularity is split. Otherwise, the $\hat{\text{I}}^*_{m+4}$ singularity is non-split.

For future purposes, we define a divisor $F=\sum_i C_i$ where $C_i$ are compact or non-compact curves carrying a singularity of type $\hat{\text{I}}^*_{n_i}$.
\item $\so(m)$ gauge algebra with $(m-8)\mathsf{F}$ can be constructed in the frozen phase by a curve $C$ of self-intersection $-1$ carrying a non-split $\text{I}_{m}$ singularity such that $F\cdot C=2$.

A non-split $\text{I}_{m}$ singularity on a $-1$ curve corresponds to a stack of $m$ D7 branes intersecting two O7 planes in type IIB language. Since $F\cdot C=2$, both of these O7 planes are O7$^+$. Hence, the gauge algebra carried by $C$ is $\so(m)$.

There are a total of $m+12$ zeroes of $\tilde\Delta_C$. 20 of these come from intersections of $C$ with the two O7$^+$ planes. This is because an O7$^+$ plane corresponds to a $\hat{\text{I}}^*_{4}$ singularity over which $\Delta$ vanish to order 10. Each remaining zero carries an $\mathsf{F}$ of $\so(m)$, thus leading to a total of $(m-8)\mathsf{F}$ of $\so(m)$.

We will also sometimes use a non-split $\text{I}_{m+1}$ on $C$ to construct $\so(m)$ with $(m-8)\mathsf{F}$. This should be viewed as a non-geometric Higgsing of $\so(m+1)$ living on $\text{I}_{m+1}$ down to $\so(m)$.
\item $\su(m)$ gauge algebra with $2m\mathsf{F}$ can be constructed in the frozen phase by the following configuration of two curves $C$ and $D$
\be \label{sum}
\begin{tikzpicture}
\node (x1) at (0,0) {$1$};
\node (y1) at (0,0.5) {$\text{I}^{ns}_{2m}$};
\node (z1) at (0,-0.5) {$C$};
\node (x2) at (1.5,0) {$2$};
\node (y2) at (1.5,0.5) {$\text{I}^s_{m}$};
\node (z2) at (1.5,-0.5) {$D$};
\draw  (x1) edge (x2);
\end{tikzpicture}
\ee
where the numbers displayed over $C$ and $D$ denote the negative of their self-intersections, the edge denotes that $C\cdot D=1$, the singularity over $C$ is non-split $\text{I}_{2m}$ and the singularity over $D$ is split $\text{I}_m$. In \cite{Bhardwaj:2018jgp}, a \emph{gauge divisor} was associated to every $6d$ gauge algebra. Here the gauge divisor for $\su(m)$ is $\Sigma=2C+D$ which means that the $6d$ gauge algebra $\su(m)$ is embedded into the $8d$ gauge algebra $\su(2m)$ carried by $\text{I}_{2m}$ with embedding index 2 and the $8d$ gauge algebra $\su(m)$ carried by $\text{I}_m$ with embedding index 1. We also need $F\cdot \Sigma=2$ for consistency, which is only possible if $F\cdot C=1$ since $D$ cannot intersect any other singularity.

It is again possible to understand this construction perturbatively. Since $F\cdot C=1$, one of the O7 planes intersecting the stack of $2m$ D7 branes on $C$ is an O7$^+$ and the other is an O7$^-$ plane thus leading to an $\su(m)$ gauge algebra with embedding index 2 on $C$. A split $\text{I}_{m}$ singularity on the $-2$ curve $D$ corresponds simply to a stack of $m$ D7 branes on $D$ leading to another $\su(m)$ there. Now we can perform a non-geometric Higgsing which combines the two $\su(m)$ living on $C$ and $D$.

$\tilde\Delta_D$ has no zeroes other than those coming from the intersection with $\text{I}^{ns}_{2m}$ singularity on $C$. $\tilde\Delta_C$ has a total of $2m+12$ zeroes. 10 out of these come from the intersection with O7$^+$ and 2 of these come from the intersection with O7$^-$. Each of the remaining zeroes carry $2\mathsf{F}$ of $\su(m)$, thus leading to a total of $2m\mathsf{F}$ of $\su(m)$.

For $m=1$ and $m=0$, we obtain new constructions for $\mathsf{A}_1$ $\cN=(2,0)$ SCFT.
\item We will need another construction for $\sp(m)$ gauge algebra with $(2m+8)\mathsf{F}$ which is
\be \label{spm}
\begin{tikzpicture}
\node (x1) at (0,0) {$4$};
\node (y1) at (0,0.5) {$\hat{\text{I}}^{*}_{m+4}$};
\node (z1) at (0,-0.5) {$C$};
\node (x2) at (1.5,0) {$1$};
\node (y2) at (1.5,0.5) {$\text{I}^{ns}_{2m}$};
\node (z2) at (1.5,-0.5) {$D$};
\draw  (x1) edge (x2);
\end{tikzpicture}
\ee
with no other frozen singularity intersecting either $C$ or $D$. If a curve carrying a frozen singularity appears in a gauge divisor, then its coefficient in the gauge divisor is the embedding index times an extra factor of half. Thus, the gauge divisor for this configuration is $\Sigma=\half C+D$.

To understand this construction perturbatively, notice that the other O7 plane intersecting $D$ is an O7$^-$ plane which reduces the gauge algebra on the stack of $2m$ D7 branes on $D$ to $\sp(m)$. We then combine this $\sp(m)$ with the $\sp(m)$ living on $C$. Unlike the previous case, the O7$^+$ plane carried by $C$ does not induce a further reduction of gauge algebra on $D$. This makes sense because $C$ and $D$ are part of the same gauge divisor.

$\tilde\Delta_C$ has a total of $4m+16$ zeroes out of which $2m$ come from the intersection with the $\text{I}^{ns}_{2m}$ singularity living over $D$. Each other zero carries a $\half\mathsf{F}$ of the low energy $\sp(m)$, thus leading to $(m+8)\mathsf{F}$ of $\sp(m)$ living on $C$. $\tilde\Delta_D$ has a total of $2m+12$ zeroes out of which $m+10$ come from the intersection with the $\hat{\text{I}}^{*}_{m+4}$ singularity living over $C$. Moreover, 2 other zeroes come from the intersection with the O7$^-$ plane. Each other zero carries an $\mathsf{F}$ of the low energy $\sp(m)$, thus leading to $m\mathsf{F}$ of $\sp(m)$ living on $D$. In total, we get $(2m+8)\mathsf{F}$ of $\sp(m)$.

We will also sometimes use
\be
\begin{tikzpicture}
\node (x1) at (0,0) {$4$};
\node (y1) at (0,0.5) {$\hat{\text{I}}^{*}_{m+5}$};
\node (z1) at (0,-0.5) {$C$};
\node (x2) at (1.5,0) {$1$};
\node (y2) at (1.5,0.5) {$\text{I}^{ns}_{2m+1}$};
\node (z2) at (1.5,-0.5) {$D$};
\draw  (x1) edge (x2);
\end{tikzpicture}
\ee
with $\Sigma=\half C+D$ to construct $\sp(m)$ with $(2m+8)\mathsf{F}$.

\item $\so(7)$ gauge algebra with $2\mathsf{S}$ can be constructed in the frozen phase by the configuration
\be\label{so7g}
\begin{tikzpicture}
\node (x1) at (0,0) {$1$};
\node (y1) at (0,0.5) {$\text{I}^{ns}_{8}$};
\node (z1) at (0,-0.5) {$C$};
\node (x2) at (1.5,0) {$3$};
\node (y2) at (1.5,0.5) {$\text{I}^{*ns}_{2}$};
\node (z2) at (1.5,-0.5) {$D$};
\draw  (x1) edge (x2);
\end{tikzpicture}
\ee
with gauge divisor $\Sigma=2C+D$ and $F\cdot C=1$, where we have performed a non-geometric Higgsing to reduce the algebra living over $\text{I}^{*ns}_{2}$ from $\so(11)$ to $\so(7)$.

$\tilde\Delta_C$ has a total of 20 zeroes. 8 out of these come from the $\text{I}^{*ns}_{2}$ singularity on $C$. 10 other zeroes come from an intersection with O7$^+$ plane. The remaining two zeroes each carry an $\mathsf{S}$ of $\so(7)$. We propose that the zeroes of $\tilde\Delta_D$ not coming from intersection with $\text{I}^{ns}_{8}$ do not carry any matter content.

\item We will also construct $\sp(5)$ with $18\mathsf{F}$ via
\be\label{sp5}
\begin{tikzpicture}
\node (x1) at (0,0) {$4$};
\node (y1) at (0,0.5) {$\hat{\text{I}}^{*ns}_{9}$};
\node (z1) at (0,-0.5) {$C$};
\node (x2) at (1.5,0) {$1$};
\node (y2) at (1.5,0.5) {$\text{I}^{ns}_{11}$};
\node (z2) at (1.5,-0.5) {$D_2$};
\begin{scope}[]
\node (x3) at (-1.5,0) {$1$};
\node (y3) at (-1.5,0.5) {$\text{I}^{ns}_{11}$};
\node (z3) at (-1.5,-0.5) {$D_1$};
\end{scope}
\begin{scope}[shift={(1.5,2.5)}]
\node (x4) at (-1.5,0) {$1$};
\node (y4) at (-1.5,0.5) {$\text{I}^{ns}_{11}$};
\node (z4) at (-1.5,-0.5) {$D_3$};
\end{scope}
\draw  (x1) edge (x2);
\draw  (x3) edge (x1);
\draw  (y1) edge (z4);
\end{tikzpicture}
\ee
with $\Sigma=\half C+D_1+D_2+D_3$ and no other frozen singularity intersects either $C$ or any $D_i$. Each $D_i$ carries $6\mathsf{F}$ situated at 6 zeroes of residual discriminant on $D_i$.
\een

\subsubsection{A new ingredient}
We will also need a gauge-theoretic ingredient arising in the frozen phase that does not admit a construction in the unfrozen phase. This is $\su(m)$ with $\mathsf{S}^2+(m-8)\mathsf{F}$ and can be constructed by a curve $C$ of self-intersection $-1$ carrying an $\text{I}_m^s$ singularity with $F\cdot C=2$. Since the intersection points of $F$ with $C$ are branch points for the monodromy, to obtain a split $I_m$, $F$ must intersect $C$ tangentially at a single point.

Out of $m+12$ zeroes of $\tilde\Delta_C$, 20 come from the tangential intersection with O7$^+$. The remaining $m-8$ zeroes each carry an $\mathsf{F}$ of $\su(m)$.

\subsection{Construction of missing SCFTs}\label{CS}
In this subsection, we will show that the frozen phase allows us to construct all the missing SCFTs listed in Section \ref{MS}.
\bit
\item (\ref{S1}) can be constructed via
\be \label{CS1}

\ee
where any singularity without a number attached to it denotes a non-compact curve\footnote{We will only display non-compact curves carrying frozen singularities.} carrying that singularity. The double edge with a tiny $t$ on top of it denotes a tangential intersection between the curve carrying $\hat{\text{I}}^*_{4}$ and the curve carrying $\text{I}^s_{n_0}$.
\item (\ref{S2}) can be constructed via
\be \label{CS2}
%
\ee
\item (\ref{S3}) can be constructed via
\be \label{CS3}
%
\ee
where the dots denote an alternating chain of \raisebox{-.125\height}{ %
}.
\item (\ref{S4}) can be constructed via
\be \label{CS4}
%
\ee
where the dots denote an alternating chain of \raisebox{-.125\height}{ %
}.
\item (\ref{S5}) can be constructed via
\be \label{CS5}
%
\ee
This shows that (\ref{S5}) exists even though it does not have any LST parent, thus demonstrating the existence of such SCFTs.
\item (\ref{S6}) can be constructed via
\be \label{CS6}
%
\ee
where the dots denote an alternating chain of \raisebox{-.125\height}{ %
}.
\item (\ref{S7}) can be constructed via
\be \label{CS7}
%
\ee
where the dots denote an alternating chain of \raisebox{-.125\height}{ %
}.
\item (\ref{S8}) can be constructed via
\be \label{CS8}
%
\ee
\item (\ref{S9}) can be constructed via
\be \label{CS9}
%
\ee
This shows that (\ref{S9}) exists even though it does not have any LST parent.
\item (\ref{S10}) can be constructed via
\be \label{CS10}
%
\ee
This shows that (\ref{S10}) exists even though it does not have any LST parent.
\eit

\subsection{Construction of missing LSTs}\label{CL}
In this subsection, we will show that the frozen phase allows us to construct all the missing LSTs listed in Section \ref{ML}.
\bit
\item (\ref{L1}) can be constructed via
\be \label{CL1}
%
\ee
We substitute $m=0$ in (\ref{CL1}) to obtain the construction for (\ref{L2}).
\item (\ref{L3}) can be constructed via
\be \label{CL2}
%
\ee
The following limit of (\ref{CL2})
\be
%
\ee
that is dual to the construction provided in \cite{Bhardwaj:2015oru} using the unfrozen phase of F-theory. Notice that the construction of \cite{Bhardwaj:2015oru} requires $\su(m)$ to be realized on a singular curve in $B$, whereas our construction realizes $\su(m)$ on a smooth curve in $B$.
\item (\ref{L4}) can be constructed via
\be \label{CL3}
%
\ee
where the $\text{I}^{s}_{6}$ is tuned to give rise to a $\half\mathsf{\Lambda}^3$.

The following limit of (\ref{CL3})
\be
%
\ee
that is dual to the construction provided in \cite{Bhardwaj:2015oru} using the unfrozen phase of F-theory. Again, notice that the construction of \cite{Bhardwaj:2015oru} requires $\su(6)$ to be realized on a singular curve in $B$, whereas our construction realizes $\su(6)$ on a smooth curve in $B$.
\item (\ref{L5}) can be constructed via
\be \label{CL4}
%
\ee
We substitute $m=0$ in (\ref{CL4}) to obtain the constructions for (\ref{L6}) and (\ref{L7}).
\item (\ref{L8}) can be constructed via
\be \label{CL5}
%
\ee
\item (\ref{L9}) and (\ref{L10}) can be constructed via
\be \label{CL6}
%
\ee
where the $\text{I}^{s}_{6}$ is tuned to give rise to a $\half\mathsf{\Lambda}^3$.
\item (\ref{L11}) can be constructed via
\be \label{CL7}
%
\ee
where the dots denote an alternating chain of \raisebox{-.125\height}{ %
} and the dashed ellipse encircling the first two curves indicates that those two curves give rise to a single gauge algebra in $6d$, which in this case is $\su(n_0)$ as we know from (\ref{sum}). Here $m_{2i}=2n_{2i}$ and $m_{2i-1}=\frac{n_{2i-1}}{2}-4$.
\item (\ref{L12}) for $j<k$ can be constructed via
\be \label{CL8}
%
\ee
where the dots denote an alternating chain of \raisebox{-.125\height}{ %
}. Here $m_{2i}=2n_{2i}$ and $m_{2i-1}=\frac{n_{2i-1}}{2}-4$ with $\text{I}^*_{m_{2i-1}}$ singularity being split for $1\le i\le j$, and $m_{2i}=2n_{2i}+1$, $m_{2i-1}=\frac{n_{2i-1}+1}{2}-4$ with $\text{I}^*_{m_{2i-1}}$ singularity being non-split for $j+1\le i\le k$. It is known \cite{Heckman:2015bfa} that the intersection of type II singularity with $\text{I}_3^{ns}=\text{I}_{m_{2k}}^{ns}$ captures a $\half\mathsf{F}$ of $\sp(1)=\sp(n_{2k})$ as required. The $\half\mathsf{F}$ of $\sp(n_{2j})$ is localized at the intersection of $\text{I}^{ns}_{m_{2j}}$ and $\text{I}^{*ns}_{m_{2j+1}}$.

(\ref{L12}) for $j=k$ can be constructed via
\be \label{CL8'}
\begin{tikzpicture}
\begin{scope}[shift={(1.5,-1.3)}]
\node (v0) at (-0.5,0.45) {$\hat{\text{I}}^*_{4}$};
\end{scope}
\node (v1) at (-0.5,0.45) {2};
\node at (-0.5,0.9) {$\text{I}^s_{n_0}$};
\begin{scope}[shift={(1.5,0)}]
\node (v2) at (-0.5,0.45) {1};
\node (v7) at (-0.5,0.9) {$\text{I}^{ns}_{2n_0}$};
\end{scope}
\begin{scope}[shift={(3,0)}]
\node (v3) at (-0.5,0.45) {4};
\node at (-0.5,0.9) {$\text{I}^{*s}_{m_1}$};
\end{scope}
\begin{scope}[shift={(4.5,0)}]
\node (v4) at (-0.5,0.45) {1};
\node at (-0.5,0.9) {$\text{I}^{ns}_{m_2}$};
\end{scope}
\begin{scope}[shift={(6,0)}]
\node (v5) at (-0.5,0.45) {4};
\node at (-0.5,0.9) {$\text{I}^{*s}_{m_{3}}$};
\end{scope}
\begin{scope}[shift={(7.5,0)}]
\node (v6) at (-0.5,0.45) {$\cdots$};
\end{scope}
\begin{scope}[shift={(9,0)}]
\node (v8) at (-0.5,0.45) {1};
\node at (-0.5,0.9) {$\text{I}_{m_{2k}}$};
\end{scope}
\begin{scope}[shift={(10.5,0)}]
\node (v9) at (-0.5,0.45) {2};
\node at (-0.5,0.9) {$\text{I}_{1}$};
\end{scope}
\draw [dashed] (0.3,0.6) ellipse (1.4 and 1);
\draw  (v1) -- (v2);
\draw  (v2) -- (v3);
\draw  (v3) -- (v4);
\draw  (v4) -- (v5);
\draw  (v5) -- (v6);
\draw  (v0) edge (v2);
\draw  (v6) edge (v8);
\draw  (v8) edge (v9);
\end{tikzpicture}
\ee
where the dots denote an alternating chain of \raisebox{-.125\height}{ \begin{tikzpicture}
\node at (-0.5,0.45) {4};
\node at (-0.5,0.9) {$\text{I}^{*s}_{m_i}$};
\end{tikzpicture}} and \raisebox{-.125\height}{ \begin{tikzpicture}
\node at (-0.5,0.45) {1};
\node at (-0.5,0.9) {$\text{I}^{ns}_{m_{i+1}}$};
\end{tikzpicture}}. Here $m_{2i}=2n_{2i}$ and $m_{2i-1}=\frac{n_{2i-1}}{2}-4$. It is well-known that the intersection of $\text{I}_1$ with $\text{I}_2=\text{I}_{m_{2k}}$ captures a full $\mathsf{F}$ of $\sp(1)=\sp(n_{2k})$, as required.

We substitute $j=0$ in (\ref{CL8}) to obtain the construction for (\ref{L12'}). Here $m_{2i}=2n_{2i}+1$, $m_{2i-1}=\frac{n_{2i-1}+1}{2}-4$ with every $\text{I}^*_{m_{2i-1}}$ singularity being non-split.
\item (\ref{L13}) can be constructed via
\be \label{CL9}
\begin{tikzpicture}
\begin{scope}[shift={(1.5,-1.3)}]
\node (v0) at (-0.5,0.45) {$\hat{\text{I}}^*_{4}$};
\end{scope}
\node (v1) at (-0.5,0.45) {2};
\node at (-0.5,0.9) {$\text{I}^s_{n_0}$};
\begin{scope}[shift={(1.5,0)}]
\node (v2) at (-0.5,0.45) {1};
\node (v7) at (-0.5,0.9) {$\text{I}^{ns}_{2n_0}$};
\end{scope}
\begin{scope}[shift={(3,0)}]
\node (v3) at (-0.5,0.45) {4};
\node at (-0.5,0.9) {$\text{I}^{*s}_{m_1}$};
\end{scope}
\begin{scope}[shift={(4.5,0)}]
\node (v4) at (-0.5,0.45) {1};
\node at (-0.5,0.9) {$\text{I}^{ns}_{m_2}$};
\end{scope}
\begin{scope}[shift={(6,0)}]
\node (v5) at (-0.5,0.45) {4};
\node at (-0.5,0.9) {$\text{I}^{*s}_{m_{3}}$};
\end{scope}
\begin{scope}[shift={(7.5,0)}]
\node (v6) at (-0.5,0.45) {$\cdots$};
\end{scope}
\begin{scope}[shift={(9,0)}]
\node (v8) at (-0.5,0.45) {4};
\node at (-0.5,0.9) {$\text{I}^{*s}_{m_{2k+1}}$};
\end{scope}
\begin{scope}[shift={(10.5,0)}]
\node (v9) at (-0.5,0.45) {1};
\node at (-0.5,0.9) {$\text{I}_{0}$};
\end{scope}
\begin{scope}[shift={(12,0)}]
\node (v10) at (-0.5,0.45) {2};
\node at (-0.5,0.9) {$\text{I}_{0}$};
\end{scope}
\draw [dashed] (0.3,0.6) ellipse (1.4 and 1);
\draw  (v1) -- (v2);
\draw  (v2) -- (v3);
\draw  (v3) -- (v4);
\draw  (v4) -- (v5);
\draw  (v5) -- (v6);
\draw  (v0) edge (v2);
\draw  (v6) edge (v8);
\draw  (v8) edge (v9);
\draw  (v9) edge (v10);
\end{tikzpicture}
\ee
where the dots denote an alternating chain of \raisebox{-.125\height}{ \begin{tikzpicture}
\node at (-0.5,0.45) {4};
\node at (-0.5,0.9) {$\text{I}^{*s}_{m_i}$};
\end{tikzpicture}} and \raisebox{-.125\height}{ \begin{tikzpicture}
\node at (-0.5,0.45) {1};
\node at (-0.5,0.9) {$\text{I}^{ns}_{m_{i+1}}$};
\end{tikzpicture}}. Here $m_{2i}=2n_{2i}$ and $m_{2i+1}=\frac{n_{2i+1}}{2}-4$.

(\ref{L14}) can be constructed via
\be \label{CL10}
\begin{tikzpicture}
\begin{scope}[shift={(1.5,-1.3)}]
\node (v0) at (-0.5,0.45) {$\hat{\text{I}}^*_{4}$};
\end{scope}
\node (v1) at (-0.5,0.45) {2};
\node at (-0.5,0.9) {$\text{I}^s_{8}$};
\begin{scope}[shift={(1.5,0)}]
\node (v2) at (-0.5,0.45) {1};
\node (v7) at (-0.5,0.9) {$\text{I}^{ns}_{16}$};
\end{scope}
\begin{scope}[shift={(3,0)}]
\node (v3) at (-0.5,0.45) {4};
\node at (-0.5,0.9) {$\text{I}^{*s}_{4}$};
\end{scope}
\begin{scope}[shift={(4.5,0)}]
\node (v4) at (-0.5,0.45) {1};
\node at (-0.5,0.9) {$\text{I}_{0}$};
\end{scope}
\begin{scope}[shift={(6,0)}]
\node (v5) at (-0.5,0.45) {4};
\node at (-0.5,0.9) {$\text{I}_{0}$};
\end{scope}
\draw [dashed] (0.3,0.6) ellipse (1.4 and 1);
\draw  (v1) -- (v2);
\draw  (v2) -- (v3);
\draw  (v3) -- (v4);
\draw  (v4) -- (v5);
\draw  (v0) edge (v2);
\end{tikzpicture}
\ee
\item (\ref{L15}) can be constructed via
\be \label{CL11}
\begin{tikzpicture}
\begin{scope}[shift={(1.5,-1.3)}]
\node (v0) at (-0.5,0.45) {$\hat{\text{I}}^*_{4}$};
\end{scope}
\node (v1) at (-0.5,0.45) {2};
\node at (-0.5,0.9) {$\text{I}^s_{n_0}$};
\begin{scope}[shift={(1.5,0)}]
\node (v2) at (-0.5,0.45) {1};
\node (v7) at (-0.5,0.9) {$\text{I}^{ns}_{2n_0}$};
\end{scope}
\begin{scope}[shift={(3,0)}]
\node (v3) at (-0.5,0.45) {4};
\node at (-0.5,0.9) {$\text{I}^{*s}_{m_1}$};
\end{scope}
\begin{scope}[shift={(4.5,0)}]
\node (v4) at (-0.5,0.45) {1};
\node at (-0.5,0.9) {$\text{I}^{ns}_{m_2}$};
\end{scope}
\begin{scope}[shift={(6,0)}]
\node (v5) at (-0.5,0.45) {4};
\node at (-0.5,0.9) {$\text{I}^{*s}_{m_{3}}$};
\end{scope}
\begin{scope}[shift={(7.5,0)}]
\node (v6) at (-0.5,0.45) {$\cdots$};
\end{scope}
\begin{scope}[shift={(9,0)}]
\node (v8) at (-0.5,0.45) {1};
\node at (-0.5,0.9) {$\text{I}^{ns}_{m_{2k}}$};
\end{scope}
\begin{scope}[shift={(10.5,0)}]
\node (v9) at (-0.5,0.45) {2};
\node at (-0.5,0.9) {$\text{I}^{*s}_{m_{2k+1}}$};
\end{scope}
\draw [dashed] (0.3,0.6) ellipse (1.4 and 1);
\draw  (v1) -- (v2);
\draw  (v2) -- (v3);
\draw  (v3) -- (v4);
\draw  (v4) -- (v5);
\draw  (v5) -- (v6);
\draw  (v0) edge (v2);
\draw  (v6) edge (v8);
\draw  (v8) edge (v9);
\end{tikzpicture}
\ee
where the dots denote an alternating chain of \raisebox{-.125\height}{ \begin{tikzpicture}
\node at (-0.5,0.45) {4};
\node at (-0.5,0.9) {$\text{I}^{*s}_{m_i}$};
\end{tikzpicture}} and \raisebox{-.125\height}{ \begin{tikzpicture}
\node at (-0.5,0.45) {1};
\node at (-0.5,0.9) {$\text{I}^{ns}_{m_{i+1}}$};
\end{tikzpicture}}. Here $m_{2i}=2n_{2i}$ and $m_{2i+1}=\frac{n_{2i+1}}{2}-4$.

(\ref{L16}) can be constructed via
\be \label{CL12}
\begin{tikzpicture}
\begin{scope}[shift={(1.5,-1.3)}]
\node (v0) at (-0.5,0.45) {$\hat{\text{I}}^*_{4}$};
\end{scope}
\node (v1) at (-0.5,0.45) {2};
\node at (-0.5,0.9) {$\text{I}^s_{6}$};
\begin{scope}[shift={(1.5,0)}]
\node (v2) at (-0.5,0.45) {1};
\node (v7) at (-0.5,0.9) {$\text{I}^{ns}_{12}$};
\end{scope}
\begin{scope}[shift={(3,0)}]
\node (v3) at (-0.5,0.45) {2};
\node at (-0.5,0.9) {$\text{I}^{*s}_{2}$};
\end{scope}
\draw [dashed] (0.3,0.6) ellipse (1.4 and 1);
\draw  (v1) -- (v2);
\draw  (v2) -- (v3);
\draw  (v0) edge (v2);
\end{tikzpicture}
\ee
\item (\ref{L17}) can be constructed via
\be \label{CL13}
\begin{tikzpicture}
\begin{scope}[shift={(1.5,-1.3)}]
\node (v0) at (-0.5,0.45) {$\hat{\text{I}}^*_{4}$};
\end{scope}
\node (v1) at (-0.5,0.45) {2};
\node at (-0.5,0.9) {$\text{I}^s_{n_0}$};
\begin{scope}[shift={(1.5,0)}]
\node (v2) at (-0.5,0.45) {1};
\node (v7) at (-0.5,0.9) {$\text{I}^{ns}_{2n_0}$};
\end{scope}
\begin{scope}[shift={(3,0)}]
\node (v3) at (-0.5,0.45) {4};
\node at (-0.5,0.9) {$\text{I}^{*}_{m_1}$};
\end{scope}
\begin{scope}[shift={(4.5,0)}]
\node (v4) at (-0.5,0.45) {1};
\node at (-0.5,0.9) {$\text{I}^{ns}_{m_2}$};
\end{scope}
\begin{scope}[shift={(6,0)}]
\node (v5) at (-0.5,0.45) {4};
\node at (-0.5,0.9) {$\text{I}^{*}_{m_{3}}$};
\end{scope}
\begin{scope}[shift={(7.5,0)}]
\node (v6) at (-0.5,0.45) {$\cdots$};
\end{scope}
\begin{scope}[shift={(9,0)}]
\node (v8) at (-0.5,0.45) {1};
\node at (-0.5,0.9) {$\text{I}^{ns}_{m_{2k}}$};
\end{scope}
\begin{scope}[shift={(10.5,0)}]
\node (v9) at (-0.5,0.45) {2};
\node at (-0.5,0.9) {$\text{I}^{*}_{m_{2k+1}}$};
\end{scope}
\draw [dashed] (0.3,0.6) ellipse (1.4 and 1);
\draw  (v1) -- (v2);
\draw  (v2) -- (v3);
\draw  (v3) -- (v4);
\draw  (v4) -- (v5);
\draw  (v5) -- (v6);
\draw  (v0) edge (v2);
\draw  (v6) edge (v8);
\draw  (v8) edge (v9);
\end{tikzpicture}
\ee
where the dots denote an alternating chain of \raisebox{-.125\height}{ \begin{tikzpicture}
\node at (-0.5,0.45) {4};
\node at (-0.5,0.9) {$\text{I}^*_{m_i}$};
\end{tikzpicture}} and \raisebox{-.125\height}{ \begin{tikzpicture}
\node at (-0.5,0.45) {1};
\node at (-0.5,0.9) {$\text{I}^{ns}_{m_{i+1}}$};
\end{tikzpicture}}. Here $m_{2i}=2n_{2i}$ and $m_{2i-1}=\frac{n_{2i-1}}{2}-4$ with $\text{I}^*_{m_{2i-1}}$ singularity being split for $1\le i\le j$, and $m_{2i}=2n_{2i}+1$, $m_{2i-1}=\frac{n_{2i-1}+1}{2}-4$ with $\text{I}^*_{m_{2i-1}}$ singularity being non-split for $i\ge j+1$. The $\half\mathsf{F}$ of $\sp(n_{2j})$ is localized at the intersection of $\text{I}^{ns}_{m_{2j}}$ and $\text{I}^{*ns}_{m_{2j+1}}$.

We substitute $j=0$ in (\ref{CL13}) to obtain the construction for (\ref{L17'}). Here $m_{2i}=2n_{2i}+1$, $m_{2i-1}=\frac{n_{2i-1}+1}{2}-4$ with every $\text{I}^*_{m_{2i-1}}$ singularity being non-split. The $\mathsf{F}$ of $\su(n_{0})$ is localized at the intersection of $\text{I}^{ns}_{2n_0}$ and $\text{I}^{*ns}_{m_{1}}$.

(\ref{L18}) can be constructed via
\be \label{CL14}
\begin{tikzpicture}
\begin{scope}[shift={(1.5,-1.3)}]
\node (v0) at (-0.5,0.45) {$\hat{\text{I}}^*_{4}$};
\end{scope}
\node (v1) at (-0.5,0.45) {2};
\node at (-0.5,0.9) {$\text{I}^s_{7}$};
\begin{scope}[shift={(1.5,0)}]
\node (v2) at (-0.5,0.45) {1};
\node (v7) at (-0.5,0.9) {$\text{I}^{ns}_{14}$};
\end{scope}
\begin{scope}[shift={(3,0)}]
\node (v3) at (-0.5,0.45) {2};
\node at (-0.5,0.9) {$\text{I}^{*ns}_{3}$};
\end{scope}
\draw [dashed] (0.3,0.6) ellipse (1.4 and 1);
\draw  (v1) -- (v2);
\draw  (v2) -- (v3);
\draw  (v0) edge (v2);
\end{tikzpicture}
\ee
with the $\mathsf{F}$ of $\su(7)$ being localized at the intersection of $\text{I}^{ns}_{14}$ and $\text{I}^{*ns}_{3}$.
\item (\ref{L19}) can be constructed via
\be \label{CL15}
\begin{tikzpicture}
\begin{scope}[shift={(1.5,-1.3)}]
\node (v0) at (-0.5,0.45) {$\hat{\text{I}}^*_{4}$};
\end{scope}
\node (v1) at (-0.5,0.45) {2};
\node at (-0.5,0.9) {$\text{I}^s_{n_0}$};
\begin{scope}[shift={(1.5,0)}]
\node (v2) at (-0.5,0.45) {1};
\node (v7) at (-0.5,0.9) {$\text{I}^{ns}_{2n_0}$};
\end{scope}
\begin{scope}[shift={(3,0)}]
\node (v3) at (-0.5,0.45) {4};
\node at (-0.5,0.9) {$\text{I}^{*s}_{m_1}$};
\end{scope}
\begin{scope}[shift={(4.5,0)}]
\node (v4) at (-0.5,0.45) {1};
\node at (-0.5,0.9) {$\text{I}^{ns}_{m_2}$};
\end{scope}
\begin{scope}[shift={(6,0)}]
\node (v5) at (-0.5,0.45) {4};
\node at (-0.5,0.9) {$\text{I}^{*s}_{m_{3}}$};
\end{scope}
\begin{scope}[shift={(7.5,0)}]
\node (v6) at (-0.5,0.45) {$\cdots$};
\end{scope}
\begin{scope}[shift={(9,0)}]
\node (v8) at (-0.5,0.45) {1};
\node at (-0.5,0.9) {$\text{I}^{ns}_{m_{2k}}$};
\end{scope}
\begin{scope}[shift={(10.5,0)}]
\node (v9) at (-0.5,0.45) {2};
\node at (-0.5,0.9) {$\text{I}^{*ss}_{0}$};
\end{scope}
\draw [dashed] (0.3,0.6) ellipse (1.4 and 1);
\draw  (v1) -- (v2);
\draw  (v2) -- (v3);
\draw  (v3) -- (v4);
\draw  (v4) -- (v5);
\draw  (v5) -- (v6);
\draw  (v0) edge (v2);
\draw  (v6) edge (v8);
\draw  (v8) edge (v9);
\end{tikzpicture}
\ee
where the dots denote an alternating chain of \raisebox{-.125\height}{ \begin{tikzpicture}
\node at (-0.5,0.45) {4};
\node at (-0.5,0.9) {$\text{I}^{*s}_{m_i}$};
\end{tikzpicture}} and \raisebox{-.125\height}{ \begin{tikzpicture}
\node at (-0.5,0.45) {1};
\node at (-0.5,0.9) {$\text{I}^{ns}_{m_{i+1}}$};
\end{tikzpicture}}. Here $m_{2i}=2n_{2i}$ and $m_{2i-1}=\frac{n_{2i-1}}{2}-4$.

(\ref{L20}) can be constructed via
\be \label{CL16}
\begin{tikzpicture}
\begin{scope}[shift={(1.5,-1.3)}]
\node (v0) at (-0.5,0.45) {$\hat{\text{I}}^*_{4}$};
\end{scope}
\node (v1) at (-0.5,0.45) {2};
\node at (-0.5,0.9) {$\text{I}^s_{4}$};
\begin{scope}[shift={(1.5,0)}]
\node (v2) at (-0.5,0.45) {1};
\node (v7) at (-0.5,0.9) {$\text{I}^{ns}_{8}$};
\end{scope}
\begin{scope}[shift={(3,0)}]
\node (v3) at (-0.5,0.45) {2};
\node at (-0.5,0.9) {$\text{I}^{*ss}_{0}$};
\end{scope}
\draw [dashed] (0.3,0.6) ellipse (1.4 and 1);
\draw  (v1) -- (v2);
\draw  (v2) -- (v3);
\draw  (v0) edge (v2);
\end{tikzpicture}
\ee
\item (\ref{L21}) can be constructed via
\be \label{CL17}
\begin{tikzpicture}
\begin{scope}[shift={(1.5,-1.3)}]
\node (v0) at (-0.5,0.45) {$\hat{\text{I}}^*_{4}$};
\end{scope}
\node (v1) at (-0.5,0.45) {2};
\node at (-0.5,0.9) {$\text{I}^s_{n_0}$};
\begin{scope}[shift={(1.5,0)}]
\node (v2) at (-0.5,0.45) {1};
\node (v7) at (-0.5,0.9) {$\text{I}^{ns}_{2n_0}$};
\end{scope}
\begin{scope}[shift={(3,0)}]
\node (v3) at (-0.5,0.45) {4};
\node at (-0.5,0.9) {$\text{I}^{*}_{m_1}$};
\end{scope}
\begin{scope}[shift={(4.5,0)}]
\node (v4) at (-0.5,0.45) {1};
\node at (-0.5,0.9) {$\text{I}^{ns}_{m_2}$};
\end{scope}
\begin{scope}[shift={(6,0)}]
\node (v5) at (-0.5,0.45) {4};
\node at (-0.5,0.9) {$\text{I}^{*}_{m_{3}}$};
\end{scope}
\begin{scope}[shift={(7.5,0)}]
\node (v6) at (-0.5,0.45) {$\cdots$};
\end{scope}
\begin{scope}[shift={(9,0)}]
\node (v8) at (-0.5,0.45) {1};
\node at (-0.5,0.9) {$\text{I}^{ns}_{m_{2k}}$};
\end{scope}
\begin{scope}[shift={(10.5,0)}]
\node (v9) at (-0.5,0.45) {2};
\node at (-0.5,0.9) {$\text{I}^{*ns}_{0}$};
\end{scope}
\draw [dashed] (0.3,0.6) ellipse (1.4 and 1);
\draw  (v1) -- (v2);
\draw  (v2) -- (v3);
\draw  (v3) -- (v4);
\draw  (v4) -- (v5);
\draw  (v5) -- (v6);
\draw  (v0) edge (v2);
\draw  (v6) edge (v8);
\draw  (v8) edge (v9);
\end{tikzpicture}
\ee
where the dots denote an alternating chain of \raisebox{-.125\height}{ \begin{tikzpicture}
\node at (-0.5,0.45) {4};
\node at (-0.5,0.9) {$\text{I}^*_{m_i}$};
\end{tikzpicture}} and \raisebox{-.125\height}{ \begin{tikzpicture}
\node at (-0.5,0.45) {1};
\node at (-0.5,0.9) {$\text{I}^{ns}_{m_{i+1}}$};
\end{tikzpicture}}. Here $m_{2i}=2n_{2i}$ and $m_{2i-1}=\frac{n_{2i-1}}{2}-4$ with $\text{I}^*_{m_{2i-1}}$ singularity being split for $1\le i\le j$, and $m_{2i}=2n_{2i}+1$, $m_{2i-1}=\frac{n_{2i-1}+1}{2}-4$ with $\text{I}^*_{m_{2i-1}}$ singularity being non-split for $j+1\le i\le k$. The $\half\mathsf{F}$ of $\sp(n_{2j})$ is localized at the intersection of $\text{I}^{ns}_{m_{2j}}$ and $\text{I}^{*ns}_{m_{2j+1}}$ where $\text{I}^{*ns}_{m_{2k+1}}:=\text{I}^{*ns}_0$.

We substitute $j=0$ in (\ref{CL17}) to obtain the construction for (\ref{L22}). Here $m_{2i}=2n_{2i}+1$, $m_{2i-1}=\frac{n_{2i-1}+1}{2}-4$ with every $\text{I}^*_{m_{2i-1}}$ singularity being non-split. The $\mathsf{F}$ of $\su(n_{0})$ is localized at the intersection of $\text{I}^{ns}_{2n_0}$ and $\text{I}^{*ns}_{m_{1}}$.

(\ref{L23}) can be constructed via
\be \label{CL19}
\begin{tikzpicture}
\begin{scope}[shift={(1.5,-1.3)}]
\node (v0) at (-0.5,0.45) {$\hat{\text{I}}^*_{4}$};
\end{scope}
\node (v1) at (-0.5,0.45) {2};
\node at (-0.5,0.9) {$\text{I}^s_{4}$};
\begin{scope}[shift={(1.5,0)}]
\node (v2) at (-0.5,0.45) {1};
\node (v7) at (-0.5,0.9) {$\text{I}^{ns}_{8}$};
\end{scope}
\begin{scope}[shift={(3,0)}]
\node (v3) at (-0.5,0.45) {2};
\node at (-0.5,0.9) {$\text{I}^{*ns}_{0}$};
\end{scope}
\draw [dashed] (0.3,0.6) ellipse (1.4 and 1);
\draw  (v1) -- (v2);
\draw  (v2) -- (v3);
\draw  (v0) edge (v2);
\end{tikzpicture}
\ee
with the $\mathsf{F}$ of $\su(4)$ being localized at the intersection of $\text{I}^{ns}_{8}$ and $\text{I}^{*ns}_{0}$.
\item (\ref{L24}) can be constructed via
\be \label{CL20}
\begin{tikzpicture}
\begin{scope}[shift={(1.5,-1.3)}]
\node (v0) at (-0.5,0.45) {$\hat{\text{I}}^*_{4}$};
\end{scope}
\node (v1) at (-0.5,0.45) {2};
\node at (-0.5,0.9) {$\text{I}^s_{n_0}$};
\begin{scope}[shift={(1.5,0)}]
\node (v2) at (-0.5,0.45) {1};
\node (v7) at (-0.5,0.9) {$\text{I}^{ns}_{2n_0}$};
\end{scope}
\begin{scope}[shift={(3,0)}]
\node (v3) at (-0.5,0.45) {4};
\node at (-0.5,0.9) {$\text{I}^{*s}_{m_1}$};
\end{scope}
\begin{scope}[shift={(4.5,0)}]
\node (v4) at (-0.5,0.45) {1};
\node at (-0.5,0.9) {$\text{I}^{ns}_{m_2}$};
\end{scope}
\begin{scope}[shift={(6,0)}]
\node (v5) at (-0.5,0.45) {4};
\node at (-0.5,0.9) {$\text{I}^{*s}_{m_{3}}$};
\end{scope}
\begin{scope}[shift={(7.5,0)}]
\node (v6) at (-0.5,0.45) {$\cdots$};
\end{scope}
\begin{scope}[shift={(9,0)}]
\node (v8) at (-0.5,0.45) {1};
\node at (-0.5,0.9) {$\text{I}^{ns}_{m_{2k}}$};
\end{scope}
\begin{scope}[shift={(10.5,0)}]
\node (v9) at (-0.5,0.45) {4};
\node (v11) at (-0.5,0.9) {$\text{I}^{*s}_{m_{2k+1}}$};
\end{scope}
\begin{scope}[shift={(12,0)}]
\node (v10) at (-0.5,0.45) {1};
\node at (-0.5,0.9) {$\text{I}^{ns}_{2m}$};
\end{scope}
\begin{scope}[shift={(10.5,1.8)}]
\node (w9) at (-0.5,0.45) {1};
\node at (-0.5,0.9) {$\text{I}^{ns}_{2m}$};
\end{scope}
\draw [dashed] (0.3,0.6) ellipse (1.4 and 1);
\draw  (v1) -- (v2);
\draw  (v2) -- (v3);
\draw  (v3) -- (v4);
\draw  (v4) -- (v5);
\draw  (v5) -- (v6);
\draw  (v0) edge (v2);
\draw  (v6) edge (v8);
\draw  (v8) edge (v9);
\draw  (v11) edge (w9);
\draw  (v9) edge (v10);
\end{tikzpicture}
\ee
where the dots denote an alternating chain of \raisebox{-.125\height}{ \begin{tikzpicture}
\node at (-0.5,0.45) {4};
\node at (-0.5,0.9) {$\text{I}^{*s}_{m_i}$};
\end{tikzpicture}} and \raisebox{-.125\height}{ \begin{tikzpicture}
\node at (-0.5,0.45) {1};
\node at (-0.5,0.9) {$\text{I}^{ns}_{m_{i+1}}$};
\end{tikzpicture}}. Here $m_{2i}=2n_{2i}$ and $m_{2i-1}=\frac{n_{2i-1}}{2}-4$.

(\ref{L25}) can be constructed via
\be\label{CL21}
\begin{tikzpicture}
\begin{scope}[shift={(1.5,-1.3)}]
\node (v0) at (-0.5,0.45) {$\hat{\text{I}}^*_{4}$};
\end{scope}
\node (v1) at (-0.5,0.45) {2};
\node at (-0.5,0.9) {$\text{I}^s_{n_0}$};
\begin{scope}[shift={(1.5,0)}]
\node (v2) at (-0.5,0.45) {1};
\node (v7) at (-0.5,0.9) {$\text{I}^{ns}_{2n_0}$};
\end{scope}
\begin{scope}[shift={(3,0)}]
\node (v3) at (-0.5,0.45) {4};
\node (v4) at (-0.5,0.9) {$\text{I}^{*s}_{m_1}$};
\end{scope}
\begin{scope}[shift={(4.5,0)}]
\node (v10) at (-0.5,0.45) {1};
\node at (-0.5,0.9) {$\text{I}^{ns}_{2m}$};
\end{scope}
\begin{scope}[shift={(3,1.8)}]
\node (w9) at (-0.5,0.45) {1};
\node at (-0.5,0.9) {$\text{I}^{ns}_{2m}$};
\end{scope}
\draw [dashed] (0.3,0.6) ellipse (1.4 and 1);
\draw  (v1) -- (v2);
\draw  (v2) -- (v3);
\draw  (v0) edge (v2);
\draw  (v3) edge (v10);
\draw  (v4) edge (w9);
\end{tikzpicture}
\ee
where $m_{1}=\frac{n_{1}}{2}-4$.
\item (\ref{L26}) can be constructed via
\be \label{CL22}
\begin{tikzpicture}
\begin{scope}[shift={(3,3.5)}]
\node (v0) at (-0.5,0.45) {$\hat{\text{I}}^*_{4}$};
\end{scope}
\node (v1) at (-0.5,0.45) {$\hat{\text{I}}^*_{4}$};
\begin{scope}[shift={(1.5,0)}]
\node (v2) at (-0.5,0.45) {1};
\node at (-0.5,0.9) {$\text{I}^{ns}_{n}$};
\end{scope}
\begin{scope}[shift={(3,1.8)}]
\node (w2) at (-0.5,0.45) {1};
\node (v10) at (-0.5,0.9) {$\text{I}^{ns}_{n}$};
\end{scope}
\begin{scope}[shift={(3,0)}]
\node (v3) at (-0.5,0.45) {4};
\node (v7) at (-0.5,0.95) {$\hat{\text{I}}^*_{n_0+4}$};
\end{scope}
\begin{scope}[shift={(4.5,0)}]
\node (v4) at (-0.5,0.45) {1};
\node at (-0.5,0.9) {$\text{I}^{ns}_{2n_0}$};
\end{scope}
\begin{scope}[shift={(6,0)}]
\node (v5) at (-0.5,0.45) {4};
\node at (-0.5,0.9) {$\text{I}^{*s}_{m_1}$};
\end{scope}
\begin{scope}[shift={(7.5,0)}]
\node (v6) at (-0.5,0.45) {$\cdots$};
\end{scope}
\begin{scope}[shift={(9,0)}]
\node (v8) at (-0.5,0.45) {1};
\node at (-0.5,0.9) {$\text{I}^{ns}_{m_{2k}}$};
\end{scope}
\begin{scope}[shift={(10.5,0)}]
\node (v9) at (-0.5,0.45) {2};
\node at (-0.5,0.9) {$\text{I}^{s}_{m}$};
\end{scope}
\draw [dashed] (3.15,0.6) ellipse (1.5 and 1);
\draw  (v1) -- (v2);
\draw  (v2) -- (v3);
\draw  (v3) -- (v4);
\draw  (v4) -- (v5);
\draw  (v5) -- (v6);
\draw  (v6) edge (v8);
\draw  (v8) edge (v9);
\draw  (w2) edge (v7);
\draw  (v0) edge (v10);
\end{tikzpicture}
\ee
where the dots denote an alternating chain of \raisebox{-.125\height}{ \begin{tikzpicture}
\node at (-0.5,0.45) {4};
\node at (-0.5,0.9) {$\text{I}^{*s}_{m_i}$};
\end{tikzpicture}} and \raisebox{-.125\height}{ \begin{tikzpicture}
\node at (-0.5,0.45) {1};
\node at (-0.5,0.9) {$\text{I}^{ns}_{m_{i+1}}$};
\end{tikzpicture}} and the dashed ellipse encircling the first two curves indicates that those two curves give rise to a single gauge algebra in $6d$, which in this case is $\sp(n_0)$ as we know from (\ref{spm}). Here $m_{2i}=2n_{2i}$ and $m_{2i-1}=\frac{n_{2i-1}}{2}-4$.
\item (\ref{L27}) for $j<k$ can be constructed via
\be \label{CL23}
\begin{tikzpicture}
\begin{scope}[shift={(3,3.5)}]
\node (v0) at (-0.5,0.45) {$\hat{\text{I}}^*_{4}$};
\end{scope}
\node (v1) at (-0.5,0.45) {$\hat{\text{I}}^*_{4}$};
\begin{scope}[shift={(1.5,0)}]
\node (v2) at (-0.5,0.45) {1};
\node at (-0.5,0.9) {$\text{I}^{ns}_{n}$};
\end{scope}
\begin{scope}[shift={(3,1.8)}]
\node (w2) at (-0.5,0.45) {1};
\node (v10) at (-0.5,0.9) {$\text{I}^{ns}_{n}$};
\end{scope}
\begin{scope}[shift={(3,0)}]
\node (v3) at (-0.5,0.45) {4};
\node (v7) at (-0.5,0.95) {$\hat{\text{I}}^*_{n_0+4}$};
\end{scope}
\begin{scope}[shift={(4.5,0)}]
\node (v4) at (-0.5,0.45) {1};
\node at (-0.5,0.9) {$\text{I}^{ns}_{2n_0}$};
\end{scope}
\begin{scope}[shift={(6,0)}]
\node (v5) at (-0.5,0.45) {4};
\node at (-0.5,0.9) {$\text{I}^{*}_{m_1}$};
\end{scope}
\begin{scope}[shift={(7.5,0)}]
\node (v6) at (-0.5,0.45) {$\cdots$};
\end{scope}
\begin{scope}[shift={(9,0)}]
\node (v8) at (-0.5,0.45) {1};
\node at (-0.5,0.9) {$\text{I}^{ns}_{m_{2k}}$};
\end{scope}
\begin{scope}[shift={(10.5,0)}]
\node (v9) at (-0.5,0.45) {2};
\node at (-0.5,0.9) {$\text{II}$};
\end{scope}
\draw [dashed] (3.15,0.6) ellipse (1.5 and 1);
\draw  (v1) -- (v2);
\draw  (v2) -- (v3);
\draw  (v3) -- (v4);
\draw  (v4) -- (v5);
\draw  (v5) -- (v6);
\draw  (v6) edge (v8);
\draw  (v8) edge (v9);
\draw  (w2) edge (v7);
\draw  (v0) edge (v10);
\end{tikzpicture}
\ee
where the dots denote an alternating chain of \raisebox{-.125\height}{ \begin{tikzpicture}
\node at (-0.5,0.45) {4};
\node at (-0.5,0.9) {$\text{I}^*_{m_i}$};
\end{tikzpicture}} and \raisebox{-.125\height}{ \begin{tikzpicture}
\node at (-0.5,0.45) {1};
\node at (-0.5,0.9) {$\text{I}^{ns}_{m_{i+1}}$};
\end{tikzpicture}}. Here $m_{2i}=2n_{2i}$ and $m_{2i-1}=\frac{n_{2i-1}}{2}-4$ with $\text{I}^*_{m_{2i-1}}$ singularity being split for $1\le i\le j$, and $m_{2i}=2n_{2i}+1$, $m_{2i-1}=\frac{n_{2i-1}+1}{2}-4$ with $\text{I}^*_{m_{2i-1}}$ singularity being non-split for $j+1\le i\le k$. The $\half\mathsf{F}$ of $\sp(1)=\sp(n_{2k})$ is localized at the intersection of $\text{I}^{ns}_{m_{2k}}=\text{I}^{ns}_{3}$ and type II singularity. The $\half\mathsf{F}$ of $\sp(n_{2j})$ is localized at the intersection of $\text{I}^{ns}_{m_{2j}}$ and $\text{I}^{*ns}_{m_{2j+1}}$.

(\ref{L27}) for $j=k$ can be constructed via
\be \label{CL23'}

\ee
where the dots denote an alternating chain of \raisebox{-.125\height}{ %
}. Here $m_{2i}=2n_{2i}$ and $m_{2i-1}=\frac{n_{2i-1}}{2}-4$. The $\half\mathsf{F}+\half\mathsf{F}$ of $\sp(1)=\sp(n_{2k})$ is localized at the intersection of $\text{I}_{m_{2k}}=\text{I}_{2}$ and $\text{I}_1$.

\item (\ref{L28}) can be constructed via
\be \label{CL24}
%
\ee
where the dots denote an alternating chain of \raisebox{-.125\height}{ %
}. Here $m_{2i}=2n_{2i}+1$ and $m_{2i-1}=\frac{n_{2i-1}+1}{2}-4$. The $\half\mathsf{F}$ of $\sp(1)=\sp(n_{2k})$ is localized at the intersection of $\text{I}^{ns}_{m_{2k}}=\text{I}^{ns}_{3}$ and type $\text{II}$ singularity. The $\so(n+1)$ is realized by $\text{I}^{ns}_{n+2}$ and the $\so(n)$ is realized by $\text{I}^{ns}_{n+1}$. The curves encircled by the dashed ellipse give rise to $\sp(n_0)$.
\item (\ref{L29}) can be constructed via
\be \label{CL25}
%
\ee
where the dots denote an alternating chain of \raisebox{-.125\height}{ %
}. Here $m_{2i}=2n_{2i}$ and $m_{2i-1}=\frac{n_{2i-1}}{2}-4$.
\item (\ref{L30}) can be constructed via
\be \label{CL26}
%
\ee
where the dots denote an alternating chain of \raisebox{-.125\height}{ %
}. Here $m_{2i}=2n_{2i}$ and $m_{2i-1}=\frac{n_{2i-1}}{2}-4$.
\item (\ref{L31}) can be constructed via
\be \label{CL27}
%
\ee
where the dots denote an alternating chain of \raisebox{-.125\height}{ %
}. Here $m_{2i}=2n_{2i}$ and $m_{2i-1}=\frac{n_{2i-1}}{2}-4$ with $\text{I}^*_{m_{2i-1}}$ singularity being split for $1\le i\le j$, and $m_{2i}=2n_{2i}+1$, $m_{2i-1}=\frac{n_{2i-1}+1}{2}-4$ with $\text{I}^*_{m_{2i-1}}$ singularity being non-split for $i\ge j+1$. The $\half\mathsf{F}$ of $\sp(n_{2j})$ is localized at the intersection of $\text{I}^{ns}_{m_{2j}}$ and $\text{I}^{*ns}_{m_{2j+1}}$.
\item (\ref{L32}) can be constructed via
\be \label{CL28}
%
\ee
where the dots denote an alternating chain of \raisebox{-.125\height}{ %
}. Here $m_{2i}=2n_{2i}+1$ and $m_{2i-1}=\frac{n_{2i-1}+1}{2}-4$. The $\so(n+1)$ is realized by $\text{I}^{ns}_{n+2}$ and the $\so(n)$ is realized by $\text{I}^{ns}_{n+1}$. The curves encircled by the dashed ellipse give rise to $\sp(n_0)$.
\item (\ref{L33}) can be constructed via
\be \label{CL29}
%
\ee
where the dots denote an alternating chain of \raisebox{-.125\height}{ %
}. Here $m_{2i}=2n_{2i}$ and $m_{2i-1}=\frac{n_{2i-1}}{2}-4$.

(\ref{L34}) can be constructed via
\be \label{CL30}
%
\ee
\item (\ref{L35}) can be constructed via
\be \label{CL31}
%
\ee
where the dots denote an alternating chain of \raisebox{-.125\height}{ %
}. Here $m_{2i}=2n_{2i}$ and $m_{2i-1}=\frac{n_{2i-1}}{2}-4$ with $\text{I}^*_{m_{2i-1}}$ singularity being split for $1\le i\le j$, and $m_{2i}=2n_{2i}+1$, $m_{2i-1}=\frac{n_{2i-1}+1}{2}-4$ with $\text{I}^*_{m_{2i-1}}$ singularity being non-split for $j+1\le i\le k$. The $\half\mathsf{F}$ of $\sp(n_{2j})$ is localized at the intersection of $\text{I}^{ns}_{m_{2j}}$ and $\text{I}^{*ns}_{m_{2j+1}}$ where $\text{I}^{*ns}_{m_{2k+1}}:=\text{I}^{*ns}_0$.

(\ref{L36}) can be constructed via
\be \label{CL33}
%
\ee
with the $\half\mathsf{F}$ of $\sp(4)$ being localized at the intersection of $\text{I}^{ns}_{8}$ and $\text{I}^{*ns}_{0}$.
\item (\ref{L37}) can be constructed via
\be \label{CL34}
%
\ee
where the dots denote an alternating chain of \raisebox{-.125\height}{ %
}. Here $m_{2i}=2n_{2i}+1$ and $m_{2i-1}=\frac{n_{2i-1}+1}{2}-4$. The $\so(n+1)$ is realized by $\text{I}^{ns}_{n+2}$ and the $\so(n)$ is realized by $\text{I}^{ns}_{n+1}$. The curves encircled by the dashed ellipse give rise to $\sp(n_0)$.

(\ref{L38}) can be constructed via
\be \label{CL35}
%
\ee
The $\so(13)$ is realized by $\text{I}^{ns}_{14}$ and the $\so(12)$ is realized by $\text{I}^{ns}_{13}$.
\item (\ref{L39}) can be constructed via
\be \label{CL36}
%
\ee
where the dots denote an alternating chain of \raisebox{-.125\height}{ %
}. Here $m_{2i}=2n_{2i}$ and $m_{2i-1}=\frac{n_{2i-1}}{2}-4$.
\item (\ref{L40}) can be constructed via
\be \label{CL37}
%
\ee
\item (\ref{L42}) can be constructed via
\be \label{CL39}
%
\ee
where the curves encircled by each dashed ellipse give rise to an $\so(7)$ with $2\mathsf{S}$ as we suggested in (\ref{so7g}).
\item (\ref{L41}) can be constructed via
\be \label{CL38}
%
\ee
where the four curves encircled by the dashed circle give rise to an $\sp(5)$ with $18\mathsf{F}$ as we suggested in (\ref{sp5}).
\eit

\section*{Acknowledgements}
The author thanks Davide Gaiotto, Patrick Jefferson, Hee-Cheol Kim, Peter Merkx, Tom Rudelius, Alessandro Tomasiello and Cumrun Vafa for valuable comments and discussions. This work is supported by NSF grant PHY-1719924.

\bibliographystyle{ytphys}
\let\bbb\bibitem\def\bibitem{\itemsep4pt\bbb}
\bibliography{refd}
\end{document}